\documentclass[aps,prb,twocolumn,floats,epsfig,pdflatex]{revtex4}
\usepackage{amsmath}
\usepackage{amsfonts}
\usepackage{graphicx}
\usepackage{amssymb}
\usepackage{amsbsy}
\usepackage{color}
\usepackage{ulem}

\newcommand{\bra}[1]{\langle{#1}|}
\newcommand{\ket}[1]{|{#1}\rangle}

\newcommand{\beq}{\begin{equation}}
\newcommand{\eeq}{\end{equation}}
\newcommand{\bea}{\begin{eqnarray}}
\newcommand{\eea}{\end{eqnarray}}

\begin{document}

\title{Dynamics of the vacuum state in a periodically driven Rydberg chain}
\author{Bhaskar Mukherjee$^1$, Arnab Sen$^1$, Diptiman Sen$^2$, and
K. Sengupta$^1$}
\affiliation{$^1$School of Physical Sciences, Indian Association for the
Cultivation of Science, Kolkata 700032, India \\
$^2$Centre for High Energy Physics and Department of Physics, Indian Institute
of Science, Bengaluru 560012, India}
\date{\today}

\begin{abstract}

We study the dynamics of the periodically driven Rydberg chain starting from the state with zero Rydberg excitations (vacuum state denoted by $|0\rangle$) using a square pulse protocol in the high drive amplitude limit. We show, using exact diagonalization for finite system sizes ($L\le 26$), that the Floquet Hamiltonian of the system, within a range of drive frequencies which we chart out, hosts a set of quantum scars which have large overlap with the $|0\rangle$ state. These scars are distinct from their counterparts having high overlap with the maximal Rydberg excitation state ($|\mathbb{Z}_2\rangle$); they coexist with the latter class of scars and lead to persistent coherent oscillations of the density-density correlator starting from the $|0\rangle$ state. We also identify special drive frequencies at which the system undergoes perfect dynamic freezing and provide an analytic explanation for this phenomenon. Finally, we demonstrate that for a wide range of drive frequencies, the system reaches a steady state with sub-thermal values of the density-density correlator. The presence of such sub-thermal steady states, which are absent for dynamics starting from the $|\mathbb{Z}_2\rangle$ state, imply a weak violation of the eigenstate thermalization hypothesis in finite sized Rydberg chains distinct from that due to the scar-induced persistent oscillations reported earlier. We conjecture that in the thermodynamic limit such states may exist as pre-thermal steady states that show anomalously slow relaxation. We supplement our numerical results by deriving an analytic expression for the Floquet Hamiltonian using a Floquet perturbation theory in the high amplitude limit which provides an analytic, albeit qualitative, understanding of these phenomena at arbitrary drive frequencies. We discuss experiments which can test our theory.

\end{abstract}


\maketitle

\section{Introduction}
\label{intro}

Recent advance in experiments using ultracold atoms have led to
great progress in understanding non-equilibrium dynamics of closed
quantum systems \cite{bloch1,blochrev,dipoleexp1}. These experiments
have shed considerable light on the many-body dynamics of strongly
interacting bosons in the presence of an experimentally applied tilt
or equivalently a synthetic electric field \cite{dipoleexp1}. More
recently, similar set of experiments have been carried out on a
chain of Rydberg atoms \cite{scarref1}. Such a chain consists of an
one-dimensional (1D) array of ultracold $^{87}{\rm Rb}$ atoms which
can be excited to a metastable excited Rydberg state by application
of a suitably designed laser. Two such atoms in their excited state
experience repulsive dipolar interaction between them. The strength
of this interaction can be tuned in these experiments; in
particular, it is possible to reach a regime where the existence of
two Rydberg atoms on neighboring sites is practically forbidden. In
addition, it is also possible to tune the on-site energy for
creating a Rydberg excitation. The low energy properties of the
Rydberg chain can therefore be described by the Hamiltonian
\begin{eqnarray} H_0 &=& \sum_j [(-\Omega |g_j\rangle\langle e_j| + {\rm H.c.})
+ \Delta \hat n_j] +\sum_{ij} V_{i-j} \hat n_i \hat n_j, \nonumber\\
\label{ryd1} \end{eqnarray} where $j$ denotes the site index,
$|g\rangle = \prod_j |g_j\rangle$ is the ground state, $|e_j\rangle$
denotes the state at site $j$ with a Rydberg excitation, ${\hat
n}_j$ denotes the number operator for the Rydberg excitations, and
$V_{i-j}= V^0/|i-j|^3$ is the interaction potential between the
excited atoms. In the regime of interest, $V^0$ is chosen such that
$V_1 \gg \Omega, \Delta \gg V_{n>1}$. In this regime the interaction
term in $H_0$ can be replaced by the constraint $\hat n_j \hat
n_{j+1}=0$ for all sites. For large negative $\Delta$, the ground
state of the system corresponds to Rydberg excitations on all
alternate sites; this state is dubbed as $|\mathbb{Z}_2\rangle$
since it breaks $\mathbb{Z}_2$ symmetry (there are two such states,
and they are related to each other by a translation by one site). In
contrast, for large positive $\Delta$, the ground state is the
vacuum of Rydberg excitation and is termed as $|0\rangle$. These two
ground states are separated by an Ising quantum phase transition at
$ \Delta=-1.31 \Omega$ \cite{subir1,subir2}. Both the $|0\rangle$
and $|\mathbb{Z}_2\rangle$ states have been observed
experimentally~\cite{scarref1}. Similar observations have also been
carried out using systems of 1D ultracold bosons in synthetic
electric fields~\cite{dipoleexp1}.

The experiments in Ref.\ \onlinecite{scarref1} also studied quench
dynamics of the Rydberg atoms starting from the $|\mathbb{Z}_2\rangle$ state
and found persistent coherent oscillatory dynamics of the Rydberg
excitations when the system is allowed to evolve after a sudden
quench of $\Delta \to 0$. This behavior constitutes a violation of the
eigenstate thermalization hypothesis (ETH) which is one of the
central paradigms for understanding out-of-equilibrium dynamics of
closed non-integrable quantum systems \cite{rev1a,rev1b,rev1c,rev1d,
rev2,deutsch1,srednicki1,rigol1}. It predicts eventual
thermalization for non-equilibrium dynamics of a generic many-body
state \cite{rev2}. This hypothesis is strongly violated in certain
cases such as 1D disordered electrons in their many-body localized 
phase~\cite{mblref1,mblref2}, but was expected to hold in disorder-free 
systems. The observed weaker failure of ETH was later understood as being 
due to the presence of quantum scars in the eigenstates of $H_0$ (with 
$\Delta=0$)~\cite{scarrefqm1,scarref1,scarref2a,scarref2b,scarref2c,
scarref2d,scarref2e,scarref3a,scarref3b,scarref3c,scarref3d,scarref3e}.

These quantum scars, which have large overlap with the initial
$|\mathbb{Z}_2\rangle$ state, are eigenstates with finite energy
density but anomalously low entanglement
entropy~\cite{scarref1,scarref2b,scarref2c,scarref2d,scarref3b}.
They form an almost closed subspace in the system's Hilbert space
under the action of its Hamiltonian and lead to persistent coherent
oscillatory dynamics of correlation functions starting from initial
states that have a high overlap with scars. This provides an
observable consequence of their presence as verified in recent
experiments on quench dynamics of a chain of ultracold Rydberg atoms
\cite{scarref1}. Such scar states, having high overlap with the
$|\mathbb{Z}_2\rangle$ state (hence the name $\mathbb{Z}_2$-scar)
have been theoretically studied using a forward-scattering
approximation (FSA) which reproduces the scar-manifold via a Lanczos
iteration starting from a $|\mathbb{Z}_2 \rangle$
state\cite{scarref2a,scarref2c,scarref2d}. The effect of $\mathbb
{Z}_2$ scars on the dynamics of a periodically driven Rydberg chain has
also been studied; it was found that the drive frequency can be used
as a tuning parameter to induce transitions between ETH violating
oscillatory and ETH obeying thermal regimes \cite{scarfl1}. Such
transitions were also shown to occur for a class of noisy and
quasiperiodic drives \cite{scarfl2}.

Analogous studies on quench dynamics starting from the $|0\rangle$
state find an expected thermalization which is consistent with ETH. However,
for periodically driven chains, where the drive frequency can act as a tuning
parameter, such dynamics has not been studied so far. In this work, we carry
out such a study by using a square pulse protocol for $\Delta$
\begin{eqnarray} \Delta &=& -\Delta_0 \quad {\rm for} \quad 0\le t\le T/2,
\nonumber\\
&=& \Delta_0 ~~\quad {\rm for} \quad T/2 < t \le T, \label{protocol1}
\end{eqnarray}
where $T=2\pi/\omega_D$ is the time period of the drive and $\omega_D$ is the
drive frequency. In what follows, we shall compute the correlation function
\begin{eqnarray} C_{j \ell} &=& \langle \psi(nT)|\hat n_j \hat n_{j+\ell}|
\psi(nT) \rangle, \label{corrfn1} \end{eqnarray}
where $|\psi(nT)\rangle$ is the state of the system after $n$ drive
cycles starting from the $|0\rangle$ state. To this end, we use
exact diagonalization for finite-sized chains of length $L\le 26$.
Our numerical results will be supplemented by an analytical, albeit
qualitative, explanation of the main features of the dynamics of the
system using a perturbative Floquet Hamiltonian. We derive this
Hamiltonian using Floquet perturbation theory with $\Omega/\Delta_0$
as the perturbation parameter \cite{dsen1,thomas1}. We note that
such a derivation is distinct from the standard Magnus or $1/\omega_D$
expansion; the Floquet Hamiltonian we obtain explains the qualitative
behavior of the system at both high and low frequency limits.

The central results that we obtain from such a study are as follows.
First, we show that for a range of drive frequencies, $C_{22}$
displays coherent oscillatory dynamics and does not thermalize. Such
a behavior constitutes a weak violation of ETH and has been reported
earlier for dynamics starting from the $|\mathbb{Z}_2\rangle$ state for
both quench and periodic driving~\cite{scarref1,scarfl1}. Its origin
in the earlier known cases has been shown to be due to the presence
of quantum scars which have high overlaps with the initial
$|\mathbb{Z}_2\rangle$ state. In our work, we show that an analogous
behavior for dynamics starting from the $|0\rangle$ state originates
from the existence of a different set of scar states in the Floquet
eigenstates of the system. These scars have high overlaps with the
$|0\rangle$ state (and hence are termed as $|0\rangle$ scars) and
coexists with the $|\mathbb{Z}_2\rangle$ scars for a range of
frequencies which we chart out. We study the properties of these scars
using the FSA reformulated using a different (compared to the
$\mathbb{Z}_2$ case) decomposition of an effective Hamiltonian which
qualitatively resembles the Floquet Hamiltonian of the driven chain.
Our analysis brings out the importance of higher spin terms for the
stability of the scar-induced oscillations. Second, we identify
specific drive frequencies at which the $|0\rangle$ state barely
evolves. This constitutes an example of dynamical
freezing~\cite{adas1,pekker1} in an experimentally realizable
non-integrable many-body system. We provide an analytic
understanding of this phenomenon by using the perturbative Floquet
Hamiltonian and by performing an exact analytical calculation for
small system sizes which predicts the freezing frequency almost
exactly. Finally, we show that lowering the drive frequency from the
dynamical freezing point with the highest frequency, we find a regime
where the system reaches a steady state with sub-thermal values of
$C_{22}$. We note that such steady states provide a new route to
weak ETH violation for finite-sized chains; it does not feature
coherent persistent oscillations and has no analogs for quench or
periodic dynamics starting from the $|\mathbb{Z}_2\rangle$ initial
state. Our numerics indicates that such behavior may persist as
a prethermal phase of thermodynamically large Rydberg chains up to a
large but finite number of drive cycles.

The rest of the paper is organized as follows. In Sec.\
\ref{floquet}, we introduce the basic model which we use for our
computations and derive the Floquet Hamiltonian for the system. This
is followed by Sec.\ \ref{numerics}, where we present our numerical
results and interpret them using the Floquet Hamiltonian. Finally,
in Sec.\ \ref{diss}, we summarize our results, discuss experiments
which can test them, and conclude. Further details of our
calculations on the derivation of the analytical form of the
perturbative Floquet Hamiltonian and analytic results regarding
dynamic freezing are presented in Appendices \ref{appA} and
\ref{appB} respectively. The details of the FSA calculation is
presented in App.\ \ref{appC}.

\section{Floquet perturbation theory}
\label{floquet}

The Hamiltonian describing the properties of an ultracold Rydberg atomic chain,
given by Eq.\ \eqref{ryd1}, can be directly mapped to a simple spin
model in the regime where $V_1 \gg \Delta,\Omega \gg V_{n>1}$. In
this regime the interaction term can be replaced by a hard
constraint on Rydberg excitations on neighboring sites. Such a
mapping is achieved by writing $\hat n_j = (\sigma_j^z + 1)/2$ and
$|e_j\rangle \langle g_j| = \sigma^+_j$, where $\sigma_j^{\alpha}$
denotes spin-1/2 Pauli matrices at site $j$ for $\alpha=x,y,z$, and
$\sigma^{\pm}_j = (\sigma_j^x \pm i \sigma_y)/2$. The constraint is
implemented by a local projection operator $P_j = (1-\sigma_j^z)/2$
\cite{scarref2c,scarref2d}. The resulting spin Hamiltonian can be
written, ignoring an unimportant constant, as~\cite{scarfl1}
\begin{eqnarray} H_{\rm spin} &=& \sum_j \left(-w \tilde \sigma_j^x
+ \frac{\lambda}{2}\sigma_j^z \right), \label{hamspin1}
\end{eqnarray}
where $\tilde \sigma^{\alpha}_j = P_{j-1} \sigma^{\alpha}_j P_{j+1}$
for $\alpha=x,y,z$. It can be easily seen that $H_{\rm spin}$ may be
identified with $H$ in Eq.\ \eqref{ryd1}, with $\Omega=w$ and
$\lambda=\Delta$. We note that $H_{\rm spin}$ also constitutes a
spin representation of the dipole model introduced in Ref.\
\onlinecite{subir1} and can thus be realized in experiments involving the
tilted Bose-Hubbard model~\cite{dipoleexp1}. For $\lambda=0$, $H_{\rm spin}$
yields the PXP model which is known to host $|\mathbb{Z}_2\rangle$
scars~\cite{scarref2a,scarref2b,scarref2c,scarref2d,scarref2e}.

In what follows, we will study the periodic dynamics of this
model using a square pulse protocol
\begin{eqnarray}
\lambda (t) &=& -\lambda \quad {\rm for} \quad 0\le t\le T/2, \nonumber\\
&=& \lambda ~~\quad {\rm for} \quad T/2 < t \le T, \label{protocol2}
\end{eqnarray}
which is identical to the protocol mentioned in Eq.\
\eqref{protocol1}. We shall be interested in the correlation function
\begin{eqnarray} O_{j \ell} &=& \frac{1}{4} \langle \psi(nT)|
(1+\sigma^z_j)(1+\sigma^z_{j+\ell})|\psi(nT) \rangle \label{corrfn2}
\end{eqnarray}
with $|\psi(0)\rangle= |0\rangle$. We note that $O_{j \ell}$ is
identical to $C_{j \ell}$ (Eq.\ \eqref{corrfn1}) for Rydberg atoms.

In the rest of this section, we shall derive a perturbative Floquet
Hamiltonian for $H_{\rm spin}$ (Eq.\ \eqref{hamspin1}) driven by the
protocol given in Eq.\ \eqref{protocol2} in the high drive amplitude
limit $\lambda/w \gg 1$ but without any approximation about the
drive frequency. In doing so, we shall use the formalism developed
in Refs.\ \onlinecite{dsen1} and \onlinecite{thomas1} and will closely follow
the approach of Ref.\ \onlinecite{thomas1}.

We treat the term $H_1=-w \sum_j \tilde \sigma_j^x$ in the
Hamiltonian as a perturbation and note that for $w=0$, the exact
evolution operator for the system can be written as (here and in the
rest of this work we set $\hbar=1$)
\begin{eqnarray} U_0 (t,0) &=& e^{i \lambda t \sum_j \sigma_j^z /2} \quad
{\rm for} \quad t\le T/2, \nonumber\\
&=& e^{i \lambda (T-t) \sum_j \sigma_z^j/2} \quad {\rm for} \quad
T/2 \le t \le T. \label{u0eq} \end{eqnarray}
$U_0$ is diagonal in the eigenbasis of $\sigma_j^z$. For simplicity
of calculation, we denote $|m\rangle$ to be set of states for which
$n_{\uparrow}-n_{\downarrow} = m$, where $n_{\uparrow(\downarrow)}$
is the number of spins with spin $\uparrow(\downarrow)$. For such
states, which form a complete basis, we find that
\begin{eqnarray} \langle m|U_0(t,0)|n\rangle &=& \delta_{mn} e^{i m \lambda
t/2} \quad {\rm for} \quad t\le T/2, \label{u0exp} \\
&=& \delta_{mn} e^{i \lambda (T-t) m/2} \quad {\rm for} \quad T/2
\le t \le T, \nonumber \end{eqnarray}
with $-L \le m\le L$ for a chain of size $L$. Note that the set
$|m\rangle$ has, in general, a large degeneracy for $w=0$ since a
particular $m$ may originate from many arrangements of spins on the
sites of the lattice. However, $m=-L$ corresponds to a
non-degenerate all down-spin state. This state is denoted as
$|0\rangle$ and shall be the initial state for this study. In this
language, the $|\mathbb{Z}_2\rangle$ state, which is doubly degenerate within
the constrained Hilbert space, corresponds to $m=0$.

Next, we compute the ${\rm O} (w)$ contribution to the evolution operator
for one time period, $U(T,0)$. To this end we compute the matrix
element of
\begin{eqnarray} U_1(T,0) &=& -i \int_0^T dt H_I(t) \label{firstorderu}
\end{eqnarray}
between states $|m\rangle$ and $|n\rangle$, where $H_I(t) =
U_0^{\dagger} (t,0) H_1 U_0(t,0)$ is the perturbation Hamiltonian in
the interaction picture. A straightforward calculation leads to
\begin{eqnarray}
\langle m|U_1(T,0)|n\rangle &=& \delta_{m,n+s} \frac{2w}{\lambda s}
\left( e^{i \lambda s T/2}-1\right), \label{matel1} \end{eqnarray}
where $s=\pm 1$. Thus in the $|m\rangle$ basis, we can write
\begin{eqnarray} U_1(T,0) &=& \sum_{m} \sum_j \sum_{s_j=\pm 1} c^{(1)}_{s_j}
|m\rangle\langle m+s_j|, \nonumber\\
c_s^{(1)} &=& \frac{4iw}{\lambda} \sin\left(\lambda T/4\right) e^{i
\lambda T s/4}, \label{u1exp} \end{eqnarray}
where the additional up or down spin in $|m+s_j\rangle$ resides on
the $j^{\rm th}$ site. Next, we note that the states $|m\rangle$ and
$|m\pm 1_j\rangle$ are connected by the projected ladder operators
$\tilde \sigma_j^{\pm} = (\tilde \sigma_j^x \pm i \tilde
\sigma_j^y)/2$ as $\tilde \sigma_j^{\pm}|m\rangle=|m\pm 1_j \rangle$.
This allows us to write the first-order Floquet Hamiltonian
$H_F^{(1)} = (i/T) U_1(T,0)$ (since $U_0(T,0)=1$ here) as~\cite{scarfl1}
\begin{eqnarray}
H_F^{(1)} &=& -w \frac{\sin(\gamma)}{\gamma} \sum_j ~[ \cos(\gamma)
\tilde \sigma_j^x + \sin(\gamma) \tilde \sigma_j^y], \label{fl1} \end{eqnarray}
where $\gamma= \lambda T/4$. We find that $H_F^{(1)}$ is identical
to the PXP model up to a global rotation and a overall
renormalization coefficient $\sin (\gamma)/\gamma$. We note that Eq.\
\eqref{fl1} was derived in Ref.\ \onlinecite{scarfl1} following a
slightly different approach~\cite{dsen1} and was used to explain the
ergodic-non-ergodic transitions for dynamics starting from the
$|\mathbb{Z}_2\rangle$ state. However, the present method allows us to derive
higher order terms in $H_F$, which, as we shall see, are crucial for
explaining the dynamics starting from the $|0\rangle$ state.

The second term in $U(T,0)$ can be obtained in a similar manner by
evaluating the matrix elements of
\begin{eqnarray} U_2(T,0) &=& (-i)^2 \int_0^T dt_1 H_I(t_1) \int_0 ^{t_1} dt_2
H_I(t_2). \label{secondorderu} \end{eqnarray}
A calculation, similar to the one carried out before and detailed in
the App.\ \ref{appA} yields
\begin{eqnarray} && U_2(T,0) = \sum_{j,j'} \sum_m \sum_{s_1,s_2=\pm}
c^{(2)}_{s_1 s_2} \tilde \sigma_j^{s_1} \tilde \sigma_{j'}^{s_2},
\label{u2exp} \\
&& c^{(2)}_{\pm \pm} = (c_{\pm}^{(1)})^2/2, ~~~{\rm and}~~~ c^{(2)}_{+-} =
c^{(2)}_{-+}= c^{(1)}_{+} c^{(1)}_-/2. \nonumber \end{eqnarray} Eq.\
\eqref{u2exp} implies that $U_2(T,0)= [U_1(T,0)]^2/2$. This ensures
that the second-order contribution to the Floquet Hamiltonian,
$H_F^{(2)}=0$. In fact, as pointed out in Ref.\
\onlinecite{scarfl1}, it can be shown that this is a consequence of
the fact that $H_F$ must satisfy the anticommutation relation $\{
\prod_{j=1, \cdots, L} \sigma_j^z, H_F \} =0$; this implies that 
$H_f^{(2n)}=0$ for all integer $n$ since terms with an even number
of $\tilde{\sigma}^{+/-}$ cannot appear in $H_F$.

Finally we proceed to obtain the third order term in $H_F$. The
corresponding evolution operator is given by
\begin{eqnarray} U_3(T,0) &=& (-i)^3 \int_0^T dt_1 H_I(t_1) \int_0 ^{t_1} dt_2
H_I(t_2) \nonumber\\
&& \times \int_0^{t_2} dt_3 H_I(t_3). \label{thirdorderu} \end{eqnarray}
As shown in App.\ \ref{appA}, the matrix elements of $U_3(T,0)$ between
any two arbitrary states $|m\rangle$ and $|n\rangle$ can be obtained
after a somewhat detailed calculation. This yields
\begin{eqnarray} U_3(T,0) &=& \sum_{j,j',j"} \sum_m \sum_{s_1,s_2,s_3=\pm}
c^{(3)}_{s_1 s_2 s_3} \tilde \sigma_{j}^{s_1} \tilde \sigma_{j'}^{s_2}
\tilde \sigma_{j"}^{s_3}, \nonumber\\
c^{(3)}_{+++} &=& (c_+^{(1)})^3/6, \quad c^{(3)}_{---} = (c_-^{(1)})^3/6,
\nonumber\\
c^{(3)}_{+--} &=& \left[ e^{3i\lambda T/2} + e^{i\lambda T/2}
(3-i\lambda T) -2 (1+ e^{i\lambda T}) \right] \nonumber\\
&& \times \frac{w^3 e^{-i\lambda T}}{\lambda^3} = c^{(3)}_{--+}, \nonumber\\
c_{-+-}^{(3)} &=& (c_{-}^{(1)})^2 c_{+}^{(1)}/2 - 2c^{(3)}_{+--},
\quad c^{(3)}_{+ - +} = c^{(3) \ast}_{- + -}, \nonumber\\
c^{(3)}_{+ + -} &=& c^{(3) \ast}_{--+},\quad c^{(3)}_{-++} = c^{(3)
\ast}_{+--}. \label{u3exp} \end{eqnarray}

Next, we compute the contribution to the Floquet Hamiltonian from
Eq. \eqref{u3exp} which comes from non-zero terms in $U_3(T,0)
-[U_1(T,0)]^3/6$. First we note, from the expressions for
the $c^{(3)}_{+++}$ and $c^{(3)}_{---}$ terms in Eq.\ \eqref{u3exp},
that all non-zero contribution to $H_F^{(3)}$ must come from terms
which have at most two $\tilde \sigma^+$ or $\tilde \sigma^-$
operators acting on different sites. All terms in $U_3(T,0)$ having
three $\tilde \sigma^+$ or $\tilde \sigma^-$ operators cancel with
similar terms from $[U_1(T,0)]^3/6$. Furthermore, the class of terms
for which the sites where the spins reside are not nearest
neighboring or same sites (so that the $\tilde \sigma^{\pm}$ on
these sites commute) do not lead to non-zero terms in $H_F^{(3)}$.
The coefficients of all such terms can be rearranged so that they
exactly cancel with similar terms from $[U_1(T,0)]^3/6$. The terms
which provide non-zero coefficient to $H_F^{(3)}$ are found to be of
three types. The first involves three spin operators on neighboring
sites such that the constraint is respected, while the second
consists of three spin operators out of which two act on the same
site. The third involves three spin operators which act on the same
site. A careful analysis of these terms leads to the third order
Floquet Hamiltonian
\begin{eqnarray} H_F^{(3)} &=& \sum_{j} ~( A_0 ~[(\tilde \sigma_{j-1}^+ 
\tilde \sigma_{j+1}^+ + \tilde \sigma_{j+1}^+ \tilde \sigma_{j-1}^+) \tilde
\sigma_j^- ~-~ 6 \tilde \sigma_j^+] \nonumber \\
&& ~~~~~~~+ ~{\rm H.c.} ), \nonumber\\
A_0 &=&\left[ e^{3i\lambda T/2} +3 e^{i\lambda T/2} (1+ i \lambda T)
+2 (1-3 e^{i\lambda T}) \right] \nonumber\\
&& \times \frac{w^3 e^{-i\lambda T}}{3 i \lambda^3 T}.
\label{fl3} \end{eqnarray}
We note that the first term in $H_F^{(3)}$ involves multiple spin
operators and generates the lowest order non-PXP terms in $H_F$. The
second term of $H_F^{(3)}$ is of the same form as in $H_F^{(1)}$ and
simply leads to a ${\rm O} (w^3)$ renormalization of its coefficients.
The former set of terms will be shown to be crucial for explaining
several properties of dynamics starting from the $|0\rangle$ state
which cannot be explained by a PXP-like Floquet Hamiltonian. The
latter class of terms will be useful for an accurate determination of
the freezing frequencies which we shall discuss in the next section.

\section{Results}
\label{numerics}

In this section, we present our numerical results on the dynamics of $O_{22}$ 
using exact diagonalization. To this end, we first note that for the chosen 
protocol (Eq.\ \eqref{protocol2}), the evolution operator is given by
\begin{eqnarray} U(T,0) = e^{-i H_{\rm spin} [\lambda]T/2} e^{-i H_{\rm
spin}[-\lambda] T/2}, \label{uevola} \end{eqnarray}
and can thus be written as
\begin{eqnarray} U(T,0) &=& \sum_{\alpha \beta} e^{-i(\epsilon_{\beta}^+ +
\epsilon_{\alpha}^-)T/2} c_{\alpha \beta}^{-+} |\alpha^-\rangle
\langle \beta^+| \label{uevol1}, \end{eqnarray}
where $\epsilon_{\alpha}^{+(-)}$ and $|\alpha^{+(-)}\rangle$ are
eigenstates and eigenfunctions of $H_{\rm spin} [+(-) \lambda]$, and
$c_{\alpha \beta}^{-+} = \langle \alpha^-|\beta^+\rangle$ denotes
eigenstate overlaps between eigenstates of $H[\lambda]$ and
$H[-\lambda]$. These eigenvalues, eigenfunctions, and the overlaps
are obtained via exact diagonalization (ED) of $H_{\rm spin}[\pm
\lambda]$ for finite system sizes $L \le 26$. This also allows us to
obtain the Floquet spectrum via diagonalization of $U(T,0)$ for
$L\le 26$. Using these, we compute the spin correlation function
$O_{22} = \langle 0|(U^{\dagger} (T,0))^n
(1+\sigma_2^z)(1+\sigma_{4}^z) U^n(T,0)|0\rangle/4$ after $n$ drive
cycles. In the limit of $n\to \infty$, the system approaches its
steady state; the value of $O_{22}$ in the steady state can be
computed using a diagonal ensemble (DE)~\cite{reimann1}. Denoting
the eigenstates of $U(T,0)$ by $|\chi_n \rangle$, it is easy to see that the
DE value of the correlator is given by
\begin{equation} O_{22}^{\rm DE} = \frac{1}{4} \sum_n |\langle 0|\chi_n\rangle
|^2 \langle \chi_n|(1+\sigma_2^z)(1+\sigma_{4}^z)|\chi_n\rangle.
\label{decorr1} \end{equation} We note that ETH predicts a steady
state value $O_{22}^{\rm DE}= 1/(\varphi^2+\varphi^4) \simeq 0.106$,
where $\varphi=(\sqrt{5}+1)/2$ is the golden ratio, which equals the
infinite temperature ensemble (ITE) value of $O_{22}$ in the
constrained Hilbert space\cite{scarfl1}.

\begin{figure*}[!]
{\includegraphics*[width=\columnwidth]{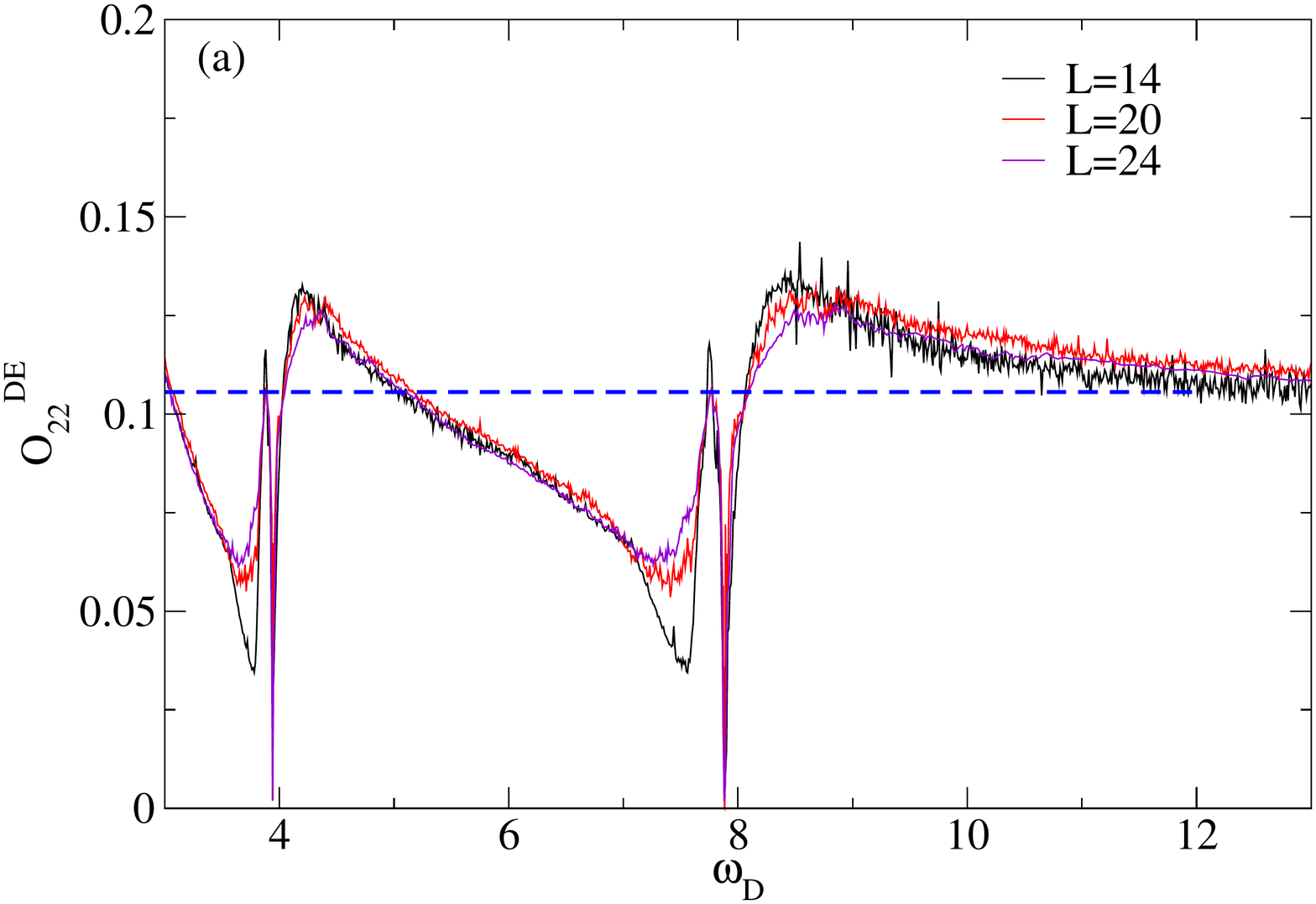}}%
{\includegraphics*[width=\columnwidth]{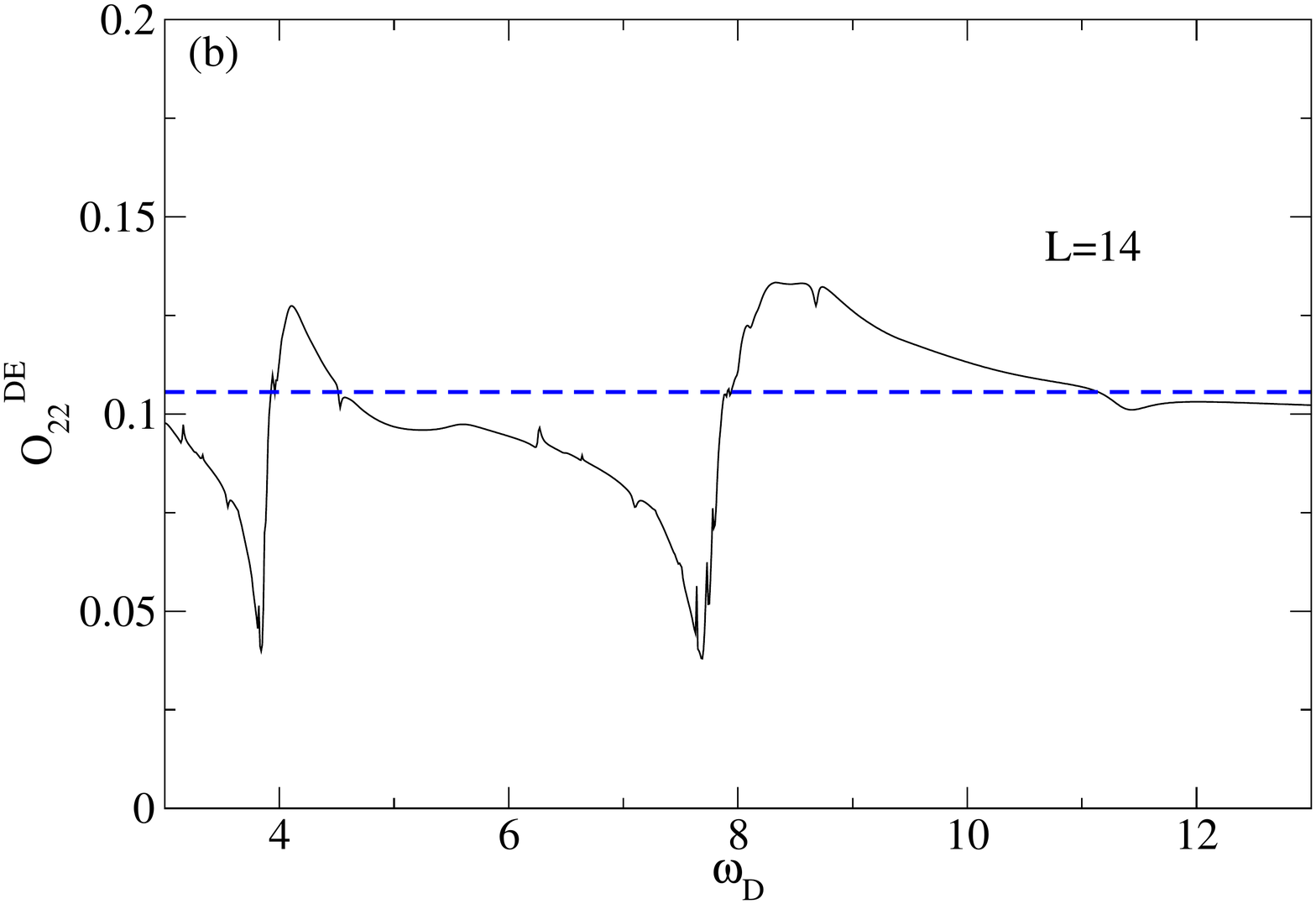}}
\caption{(a) Plots of $O_{22}^{\rm DE}$ as a
function of the drive frequency starting from an initial state $|0\rangle$ for
$L=14, 18, 24$. The blue dashed line indicates the infinite temperature
thermal value of $O_{22}^{\rm DE}$ as predicted by ETH. The plots clearly
indicate super-thermal and sub-thermal values of $O_{22}^{\rm DE}$
over a range of frequencies and dynamical freezing at specific
frequencies where $O_{22}^{\rm DE} \simeq 0$. (b) Similar plot for
$O_{22}^{DE}$ as a function of $\omega_D$ for $L=14$ as obtained
using the analytic perturbative Floquet Hamiltonian given by Eqs.\
\eqref{fl1} and \eqref{fl3}. All energies and frequencies are scaled in
units of $w/\sqrt{2}$, $\hbar=1$, and $\lambda=15$ in rescaled units
for all plots.} \label{fig1} \end{figure*}

\begin{figure}
{\includegraphics*[width=\linewidth]{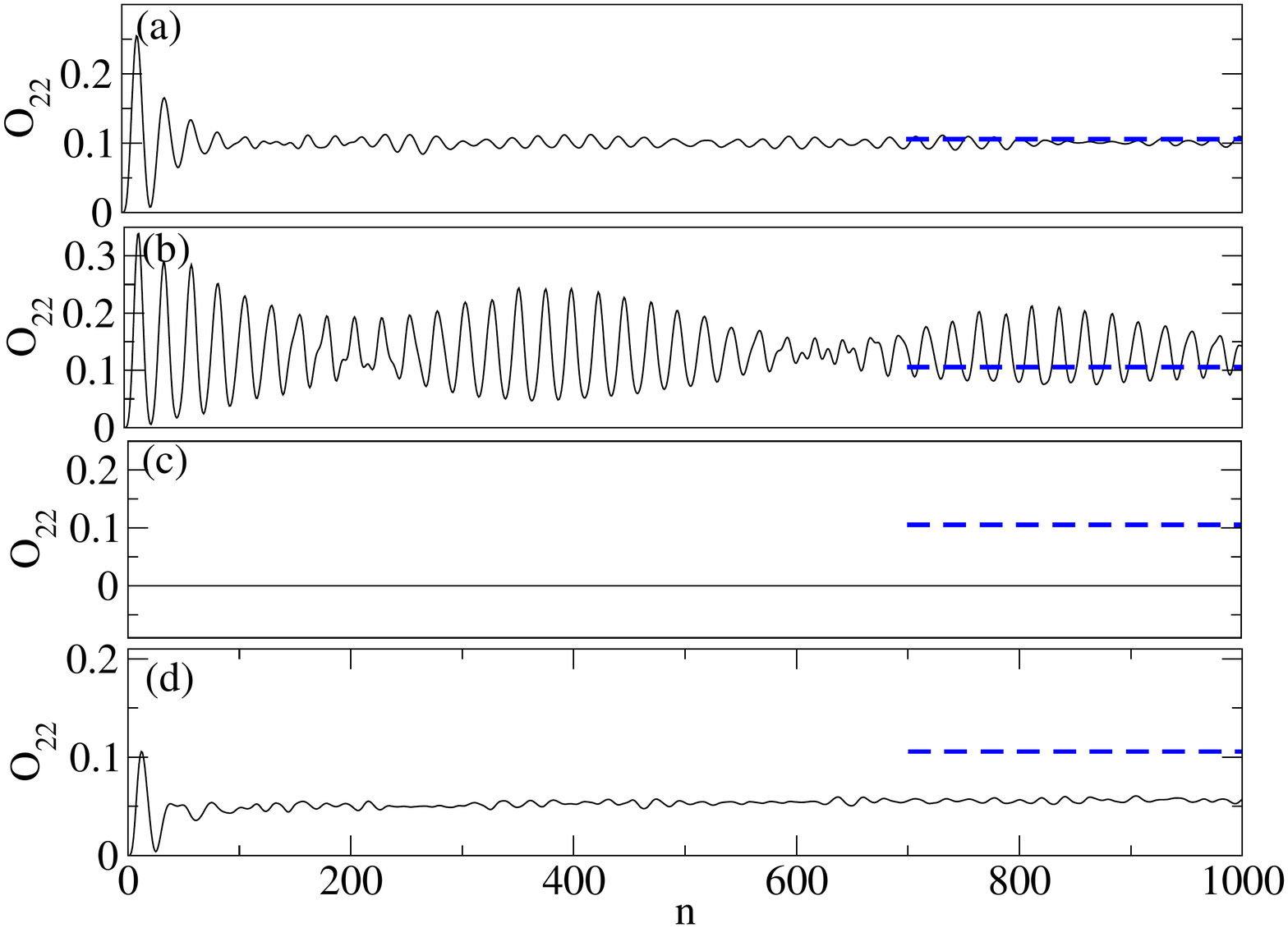}}
\caption{Plots of $O_{22}$ as a function of the number of drive cycles $n$ 
(stroboscopic time) starting from an initial state
$|0\rangle$ for (a) $\omega_D=100$, (b) $\omega_D=8.5$, (c)
$\omega_D=7.88$, and (d) $\omega_D=7.26$. The blue dashed line indicates the 
ETH predicted thermal value of $O_{22}$. Here $L=26$, $\lambda=15$, and all
units are the same as in Fig.\ \ref{fig1}.} \label{fig2a} \end{figure}

A plot of $O_{22}^{\rm DE}$, computed from the exact evolution
operator, is shown in Fig.\ \ref{fig1} (a) as a function of the
drive frequency $\omega_D$. The corresponding plot, obtained
starting from the analytical Floquet Hamiltonian at
$\mathcal{O} (w^3)$ (Eqs.\ \eqref{fl1} and \eqref{fl3}), is shown in
Fig.\ \ref{fig1} (b). From these plots, we note the following
features. First, we find that $O_{22}^{\rm DE}$ obtained using the
analytic Floquet Hamiltonian provides a qualitative match with that
obtained from exact numerics. This brings out the importance of the
multiple-spin term in Eq.\ \eqref{fl3}; the PXP Floquet Hamiltonian
(Eq.\ \eqref{fl1} and the single spin term in Eq.\ \eqref{fl3}), for
dynamics starting from the $|0\rangle$ state, predict a featureless
thermal value of $O_{22}^{\rm DE}$ as a function of $\omega_D$.
Second, we note that $O_{22}^{\rm DE}$ reaches the expected infinite
temperature thermal steady state value predicted by ETH (blue dashed
line in Fig.\ \ref{fig1}) for high frequencies. This clearly
indicates that $|\mathbb{Z}_2\rangle$ scars do not play a role in
the dynamics. In contrast, at finite $\omega_D$, there are several
non-ETH like features present as a function of the drive frequency
at least up to $L=24$ (Fig.\ \ref{fig1} (a)). Third, for $8 \le
\omega_D \le 12$, Fig.\ \ref{fig1} (a) shows that $O_{22}^{DE}$
reaches super-thermal values; this phenomenon constitutes a violation
of ETH for finite-sized chains $L \le 24$. We shall discuss this
feature in detail in Sec.\ \ref{sup1}. Fourth, for $\omega_D \simeq
7.88, 3.94 ..$, $O^{\rm DE}_{22}$ remains pinned to its initial
value ($=0$); this constitutes an example of dynamical freezing at
specific drive frequencies which we discuss in Sec.\ \ref{freez1}.
Finally, for $5 \le \omega_D \le 7.5$, we find that $O_{22}^{\rm
DE}$ exhibits sub-thermal steady-state values. This constitutes
another class of violation of ETH for finite chains which we discuss
in Sec.\ \ref{sub1}. We note that the time evolution of $O_{22}$ as
a function of the number of drive cycles, shown in Fig.\ \ref{fig2a}, in
these three regimes shows qualitatively distinct behaviors which can
be discerned in realistic experiments involving Rydberg atom chains.

\begin{figure}
{\includegraphics*[width=\linewidth]{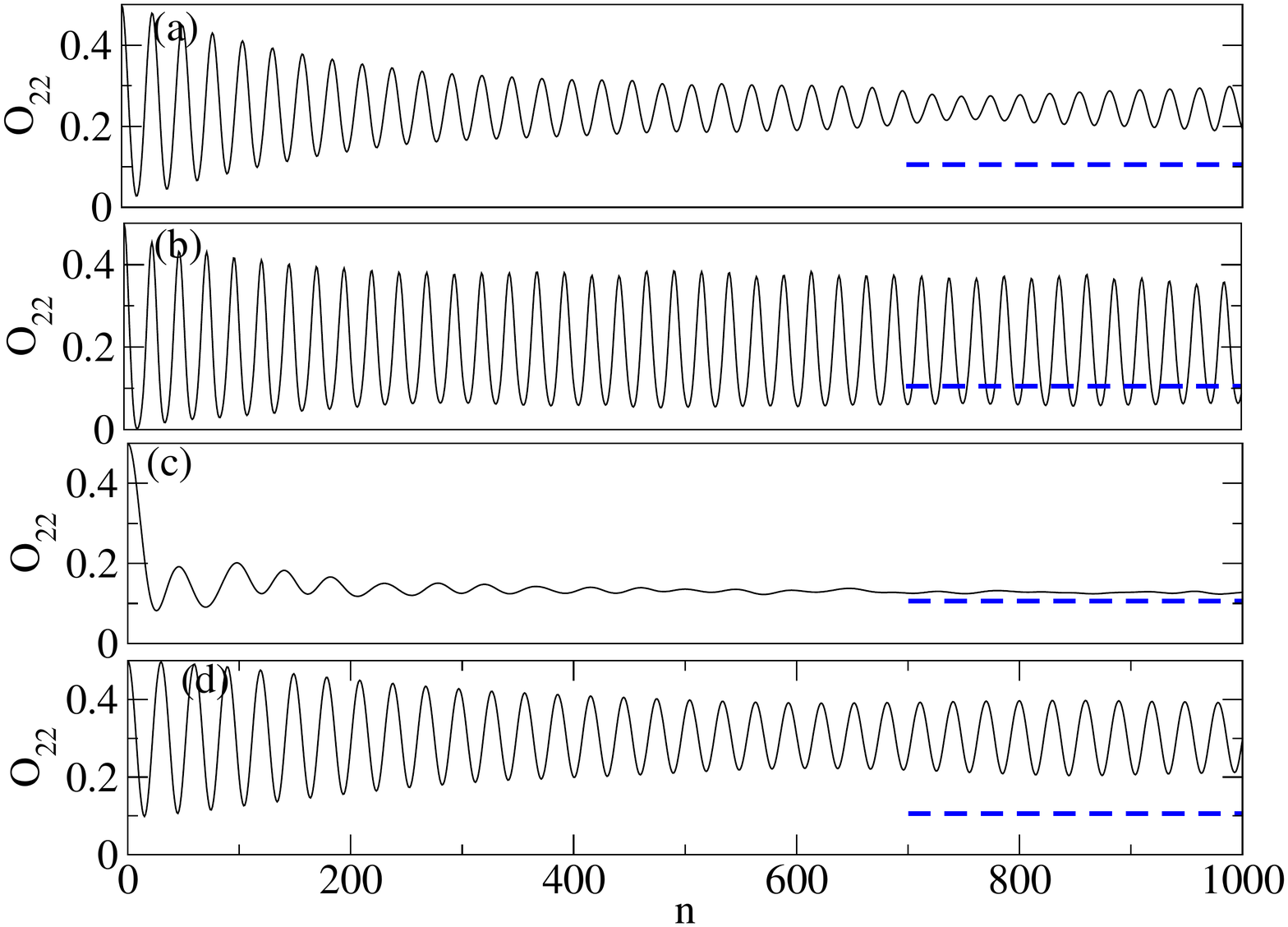}}
\caption{Plots of $O_{22}$ as a function of the number of drive cycles $n$ 
(stroboscopic time) starting from an initial state $|Z_2\rangle$ for 
(a) $\omega_D=100$, (b) $\omega_D=8.5$, (c) $\omega_D=7.88$, and 
(d) $\omega_D=7.26$. The blue dashed line indicates the ETH predicted 
thermal value of $O_{22}$. Here $L=26$, $\lambda=15$, and all
units are the same as in Fig.\ \ref{fig1}} \label{fig2b} \end{figure}

\subsection{Super-thermal steady state value}
\label{sup1}

The stroboscopic time evolution of $O_{22}$ starting from the
$|0\rangle$ state is shown in Fig.\ \ref{fig2a} for $L=26$. The
corresponding behavior of the same correlator starting from the
$|\mathbb{Z}_2\rangle$ state is shown in Fig.\ \ref{fig2b}. First,
we note that for high frequencies such as $\omega_D=100$, Fig.\
\ref{fig2a} (a) shows expected thermalization while Fig.\
\ref{fig2b} (a) shows scar-induced oscillations. This behavior is
consistent with earlier studies of the PXP
model~\cite{scarref2a,scarref2b,scarref2c,scarref2d,scarfl1} which
reported thermalization for dynamics starting from the $|0\rangle$
state in cases of both quench and periodic protocol at high drive
frequency. Panel (b) for Figs.\ \ref{fig2a} and \ref{fig2b}, in
contrast, indicate the presence of persistent oscillations for
$\omega_D=8.5$ for dynamics starting from {\it both} $|0\rangle$ and
$|\mathbb{Z}_2\rangle$ states. This leads to weak violation of ETH
and super-thermal value of $O_{22}^{\rm DE}$ for dynamics starting
from the $|0\rangle$ state.

\begin{figure}
{\includegraphics*[width=\linewidth]{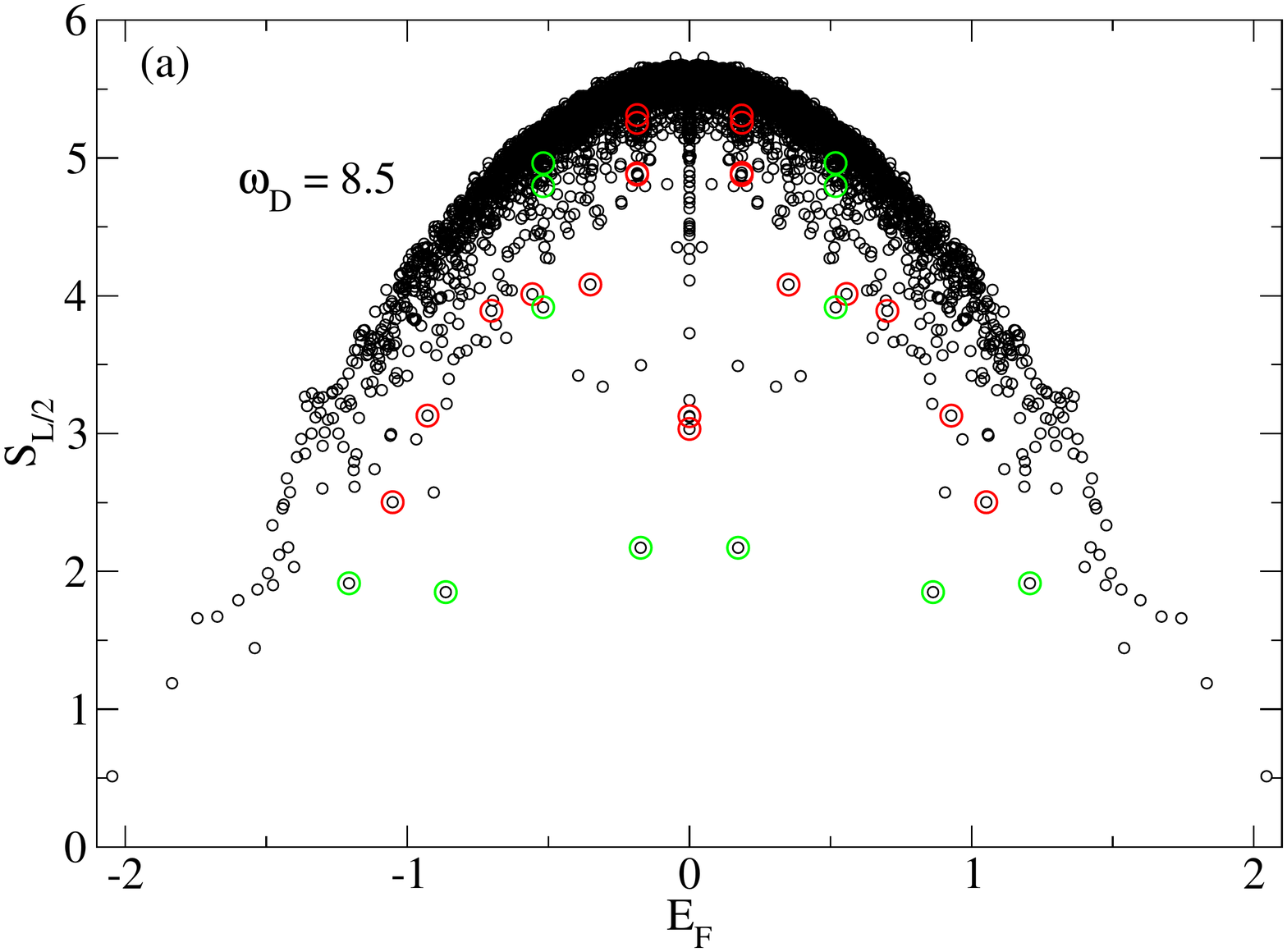}}\\
{\includegraphics*[width=\linewidth]{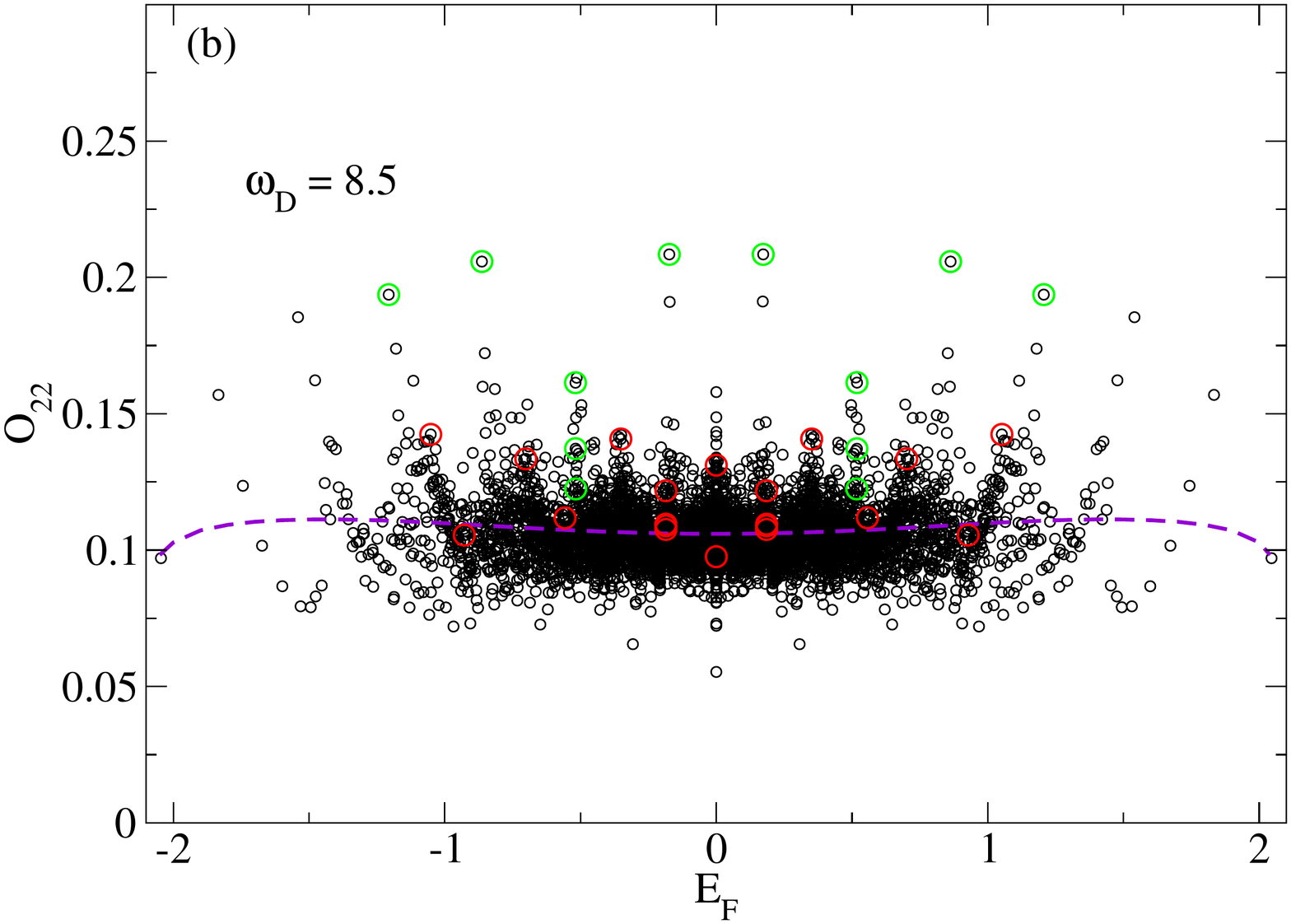}}
\caption{(a) Plot of $S_{L/2}$ for the eigenstates of $H_F$ for $L=26$ and 
$\omega_D =8.5$ at $\lambda=15$. The eigenstates with overlap $>0.01$
with $|0\rangle ~(|\mathbb{Z}_2\rangle)$ are shown using red (green) circles.
These states are distinct and coexist at this drive frequency.
(b) Plot of $O_{22}^m$ as a function of Floquet eigenstate
quasienergies $E_F$. The violet dashed line indicates the ETH
predicted value of $O_{22}$ at a temperature $T_0 (E_F)$. All
units are the same as in Fig.\ \ref{fig1}.} \label{fig3} \end{figure}

To understand the origin of these oscillations, we show the
half-chain entanglement entropy $S_{L/2}$ of the eigenstates
$|\chi_m\rangle$ of the Floquet Hamiltonian for
$\omega_D=8.5$ in Fig.\ \ref{fig3} (a).
The details of this computation have been charted out in Ref.\
\onlinecite{scarfl1}. Fig.\ \ref{fig3} (b) shows the value of
\begin{eqnarray} O_{22}^m = \frac{1}{4} \langle \chi_m|(1+\sigma_2^z)(1+
\sigma^z_4)|\chi_m\rangle \label{eigenexp} \end{eqnarray}
for all Floquet eigenstates $|\chi_m\rangle$ as a function of the
Floquet quasienergies $E_F$. The dotted line in this plot
indicates the ETH value of $O_{22}$ at a temperature
$T_0(E_F)$ as a function of these quasienergies.
Here $T_0(E_F)$ is defined such that the average
quasienergy equals $E_F$ for a canonical ensemble
with temperature $T_0(E_F)$.

\begin{figure}
{\includegraphics*[width=\linewidth]{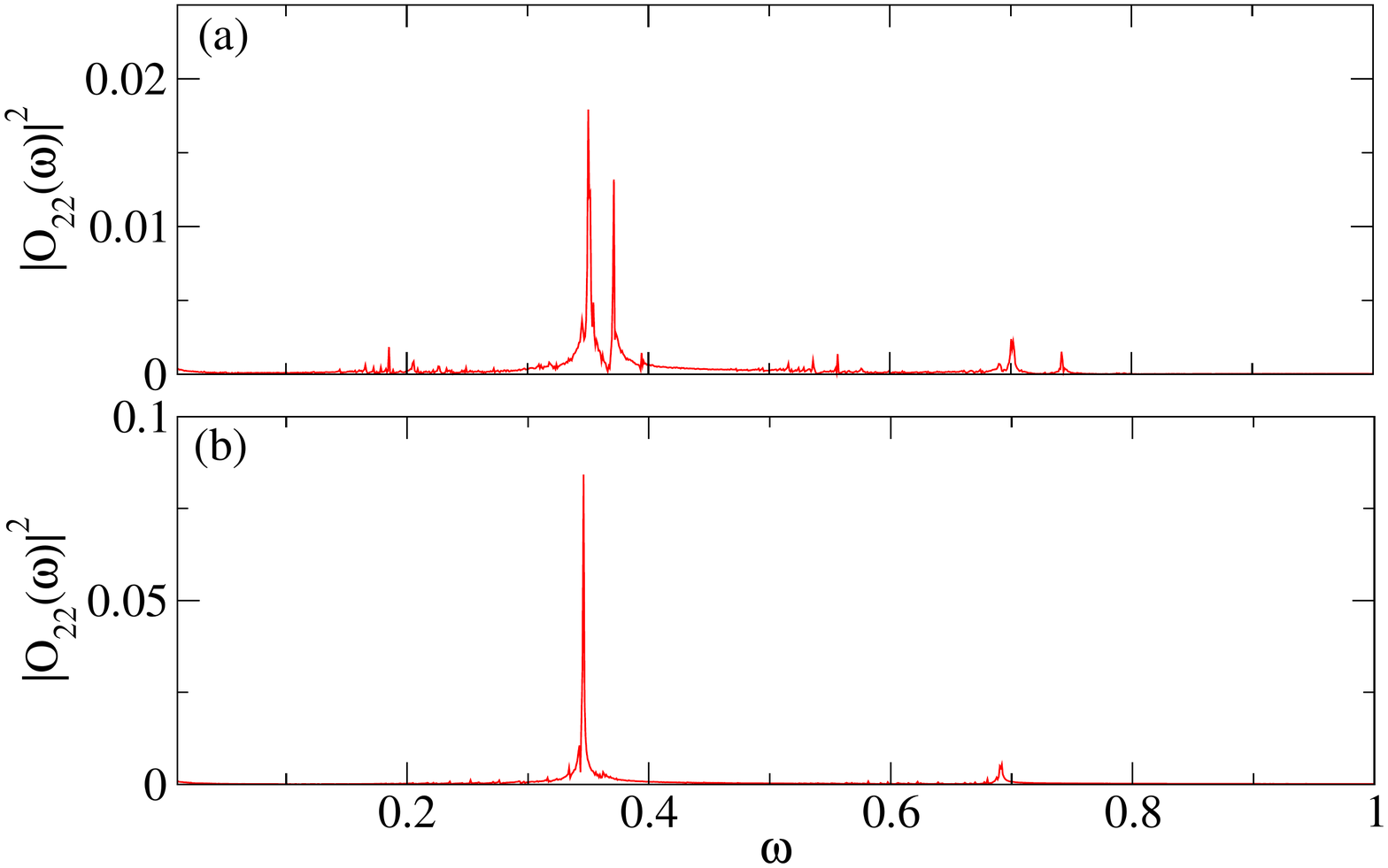}} \caption{Plots of
$|O_{22} (\omega)|^2$ obtained from the Fourier transforms of $O_{22} (n)$ 
for $L=26$, $\omega_D=8.5$ and $\lambda=15$ for the initial state being 
(a) $|0\rangle$ and (b) $|\mathbb{Z}_2 \rangle$ respectively. All units are 
the same as in Fig.\ \ref{fig1}.} \label{FT} \end{figure}

Fig.\ \ref{fig3} (a) shows the usual thermal ETH band with large
$S_{L/2}$ along with sub-thermal states with lower values of
$S_{L/2}$. The states $|\chi_m\rangle$ with $|\langle
0|\chi_m\rangle|^2 (|\langle Z_2 |\chi_m\rangle|^2)>0.01$ which
control the dynamics starting from the $|0\rangle (|Z_2\rangle)$
state is shown in red (green) circles in both panels. From Fig.\
\ref{fig3} (a), we find that the low-entropy eigenstates of $H_F$
which control the dynamics are distinct for $|0\rangle$ and
$|\mathbb{Z}_2\rangle$ initial states; at $\omega_D=8.5$, these
states coexist with each other. Furthermore, the low-entropy
eigenstates with large overlap with the $|0\rangle$ show values of
$O_{22}^m$ closer to the ETH line compared to their counterpart for
the $|\mathbb{Z}_2\rangle$ state as can be clearly seen from Fig.\
\ref{fig3} (b); thus we expect $O_{22}^{\rm DE}$ starting from the
$|0\rangle$ state to be closer to the ETH value compared to its
$|\mathbb{Z}_2\rangle$ counterpart. Nevertheless, a finite number of
these eigenstates contributing to the $|0\rangle$ dynamics are not
thermal as can be seen from Fig.\ \ref{fig3} (a). They have
significantly lower values of $S_{L/2}$ compared to the eigenstates
in the thermal band, and lead to persistent coherent oscillatory
dynamics of $O_{22}$ starting from the $|0\rangle$ state. We
therefore dub these states as $|0\rangle$ scars. Our findings
indicate that there are at least two distinct types of scars in the
Floquet spectrum of $H_{\rm spin}$ driven by the square pulse
protocol given in Eq.\ \eqref{protocol2}; this phenomenon has no
analog in the PXP model studied earlier where only
$|\mathbb{Z}_2\rangle$ scars exist. We further note that the energy
spacings between these $|0\rangle$ scar states are non-uniform
unlike their $|\mathbb{Z}_2\rangle$ counterparts; this causes a
strong beating phenomenon in the oscillation of $O_{22}$ (Fig.\
\ref{fig2a} (b)) which is much weaker for the corresponding
$|\mathbb{Z}_2\rangle$ dynamics (Fig.\ \ref{fig2b} (b)). This can be
more clearly seen in Fig.\ \ref{FT} where the Fourier transform of
$O_{22}$ starting from $|0\rangle$ (Fig.\ \ref{fig2a} (b)) and from
$|\mathbb{Z}_2\rangle$ (Fig.\ \ref{fig2b} (b)) are shown in Fig.\
\ref{FT} (a) and Fig.\ \ref{FT} (b) respectively.

As $\omega_D$ is increased, we find that the $|0\rangle$ scars merge
with the thermal band and cannot be distinguished from them for
$\omega_D>12$ where $O_{22}$ starts displaying thermal behavior
consistent with ETH (Fig.\ \ref{fig2a} (a)); in contrast, the
$|\mathbb{Z}_2\rangle$ scars persist at arbitrary high frequency
(Fig.\ \ref{fig2b} (a)). This clearly demonstrates that the
$|0\rangle$ scars require higher spin terms in $H_F$ such as the
first term of Eq.\ \eqref{fl3}; they are not eigenstates of the high
frequency Floquet Hamiltonian which constitutes a renormalized PXP
model. Finally, it is important to note here that the $|0\rangle$
scars have a higher entanglement entropy compared to their
$|\mathbb{Z}_2 \rangle$ counterparts (Fig.\ \ref{fig3} (a)) and thus
they may be more fragile to increasing system sizes.

The role of the three or higher-spin terms in $H_F$ for the
stability of $|0\rangle$ scars and the consequent coherent
oscillations can be qualitatively understood using the FSA. To this end,
we consider an effective Hamiltonian
\begin{eqnarray} H_1= -\sum_{j} \tilde \sigma^x_j + h \sum_{j} ( \tilde 
\sigma^+_{j} \tilde \sigma^-_{j-1} \tilde \sigma^-_{j+1} + {\rm H.c.}), 
\label{fsaham1} \end{eqnarray}
which qualitatively mimics $H_F$ found in Sec.\ \ref{floquet},
albeit with real valued coefficient $h$. Here we use $h$ as a tuning
parameter and study the properties of the scar-induced oscillations
within the FSA starting from $|0\rangle$. For this, we write $H_1= H^+ + H^-$ 
(with $H^-= (H^+)^\dagger$) and choose $H^- = -w \sum_j \tilde \sigma^{-}_j 
+ h \sum_j \tilde \sigma^+_{j} \tilde \sigma^-_{j-1} \tilde \sigma^-_{j+1}$
so that $H^-|0\rangle=0$. The repeated application of $H^+$ on
$|0\rangle$ (forward scattering) then generates a closed Krylov
subspace. Following standard procedure, we designate a particular
forward scattering step to be exact when the action of $H^+$ on the
Krylov vector in that step can be totally reversed by the action of
$H^-$. The inexact FSA steps generate errors which we aim to
minimize. The details of this analysis is charted out in App.\
\ref{appC}. The main results that come out of this analysis are as
follows. First, we find that in the bare PXP model ($H_1(h=0)$) all
forward scattering action are inexact after the first two FSA steps;
these errors cannot be minimized for $h=0$. This shows that the FSA
predicts instability of the $|0\rangle$ scars within the
PXP model. Second, we find that $h$ provides a control knob which
can minimize the FSA errors at different steps, although there is no
common value of $h$ for which errors in all the FSA steps are
simultaneously minimized. Our analysis finds that the errors for the
third FSA step (which is also the first error generating FSA step)
is minimized for $h^{\rm min} \simeq 0.3$; furthermore, errors in
other FSA steps are minimum close to (but not exactly at) $h=h^{\rm
min}$. The details of this procedure and the $L$ dependence of this
result is detailed out in App.\ \ref{appC}. Our analysis thus brings out
the importance of higher-spin terms in $H_1$ (and $H_F$) for
the stability of $|0\rangle$ scars. Finally, we find that the addition
of further terms such as a five-spin term to $H_1$ (see App.\
\ref{appC}) can lead to further amplification of scar-induced
oscillations and chart out the values of coefficients which achieves
such amplification. The Floquet Hamiltonian $H_F$ provides a natural
setting for generating such longer-ranged terms as the drive
frequency $\omega_D$ is lowered.

\subsection{Dynamical freezing}
\label{freez1}

\begin{figure}
{\includegraphics*[width=\linewidth]{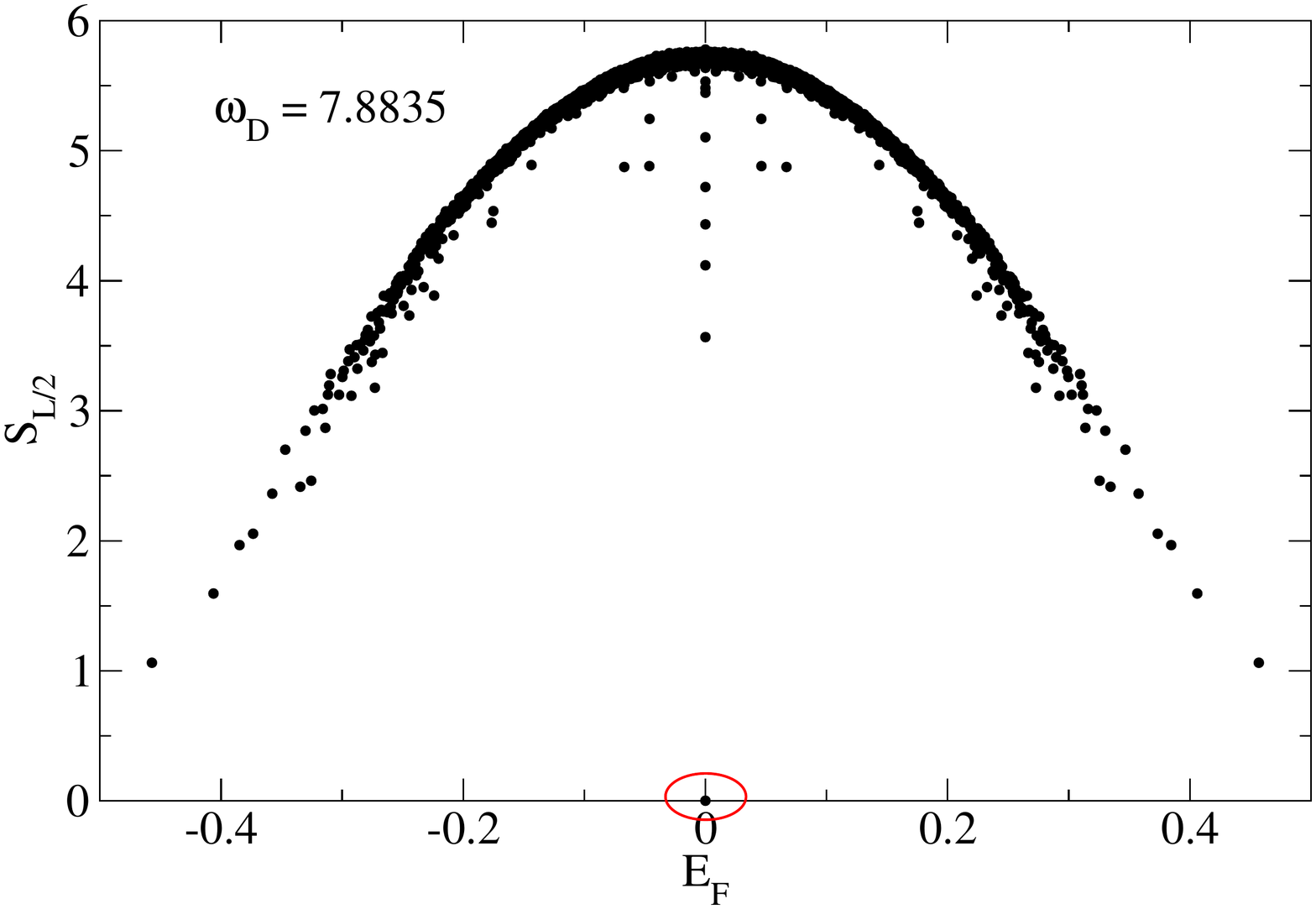}} \caption{Plot of
$S_{L/2}$ for the eigenstates of $H_F$ for $L=26$ and
$\omega_D=7.8835$ at $\lambda=15$. The state encircled in red shows
the presence of a quantum scar with a very high overlap ($>0.9999$)
with the $|0\rangle$ state. All units are the same as in Fig.\
\ref{fig1}.} \label{figsinglescar} \end{figure}

For $\omega_D=7.88$, we find from Fig.\ \ref{fig2a} (c), that the
$|0\rangle$ state, in spite of not being an eigenstate of $H_F$,
does not exhibit almost any time evolution. This phenomenon is also
found for other lower, subharmonic, drive frequencies as seen in
Fig.\ \ref{fig1} (a) where $O_{22}^{\rm DE}$ exhibits a sharp dip to
$0$. This is in sharp contrast to the evolution of the
$|\mathbb{Z}_2\rangle$ state which shows thermalization at this
frequency (Fig.\ \ref{fig2b} (c))~\cite{scarfl1}. This behavior
constitutes an example of dynamical freezing which we now discuss.
At these dynamic freezing frequencies, a quantum scar state with
vanishingly small entanglement has an almost perfect overlap ($>
0.9999$ till $L=26$) with the $|0\rangle$ state as shown by the
behavior of $S_{L/2}$ in Fig.\ \ref{figsinglescar}.

\begin{figure}
{\includegraphics*[width=\linewidth]{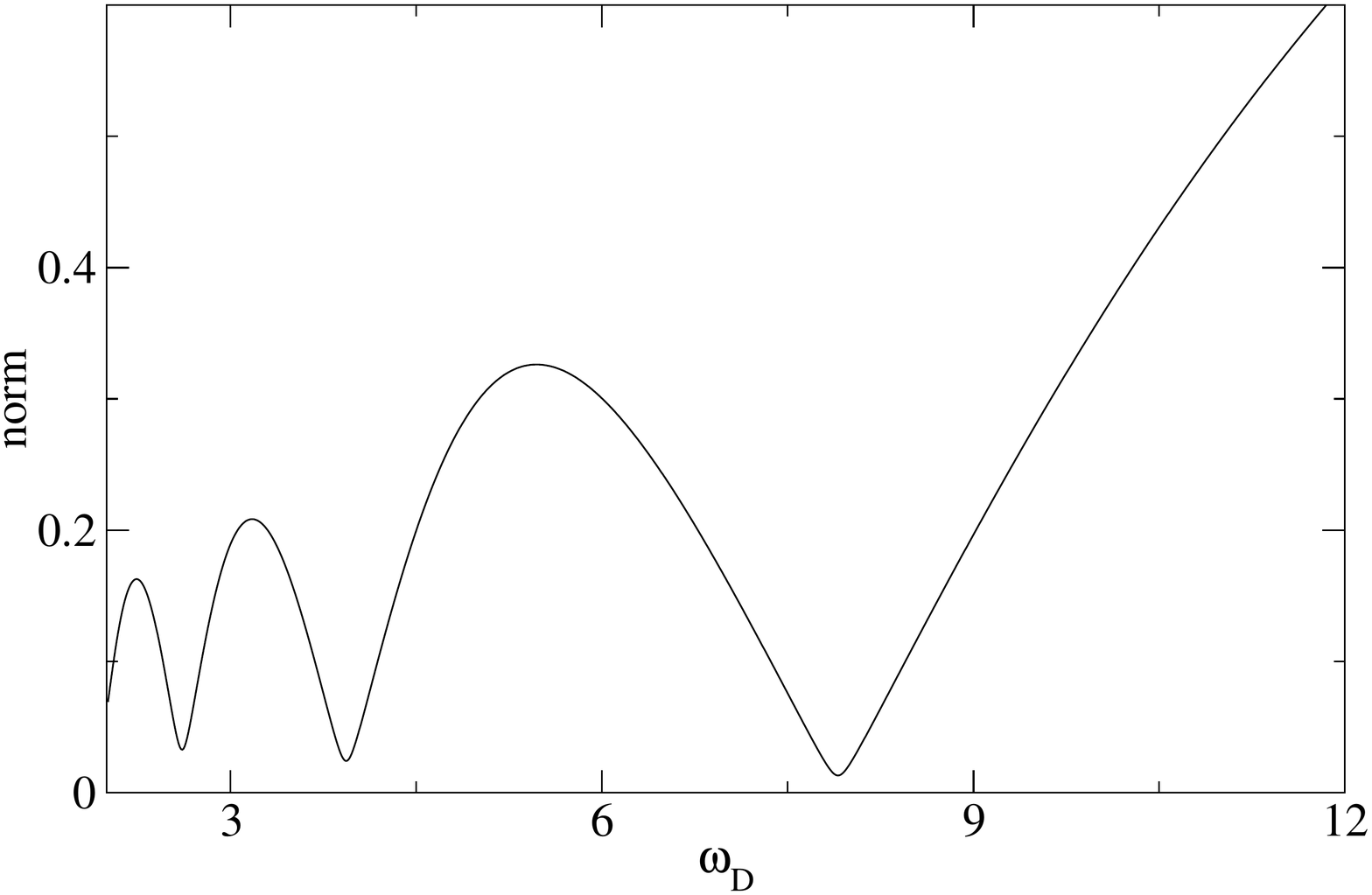}}
\caption{Plot of the norm ${\mathcal N}$ as a function of $\omega_D$. Here 
$\lambda=15$ and all units are the same as in Fig.\ \ref{fig1}. See text for 
details.} \label{fig4} \end{figure}

We first obtain a qualitative understanding of this phenomenon using
the Floquet Hamiltonian given by Eqs.\ \eqref{fl1} and \eqref{fl3}. To
this end we note that $\tilde \sigma_j^- |0\rangle = 0$ for any $j$.
Thus the first term in Eq.\ \eqref{fl3} (and any higher order terms in
$H_F$ which have $\tilde \sigma_j^-$) annihilates the state
$|0\rangle$. Consequently, the only non-trivial terms in $H_F$
contributing to the evolution of the $|0\rangle$ state are the single
spin terms charted out in Eqs.\ \eqref{fl1} and \eqref{fl3}. These single
spin terms can be written as
\begin{eqnarray} H_F^{\rm single} &=& -\sum_j \Big[\left(w
\frac{\sin(2\gamma)}{2\gamma} +
6 {\rm Re}[A_0] \right) \tilde \sigma_j^x \nonumber\\
&& \quad \quad \quad + \left(w \frac{\sin^2 \gamma}{\gamma} - 6 {\rm Im}[A_0]
\right) \tilde \sigma^y_j\Big] \nonumber\\
&=& \sum_j ~(C_1 \tilde \sigma_j^x + C_2 \tilde \sigma_j^y). \end{eqnarray}
For drive frequencies where the norm ${\mathcal N}=
\sqrt{C_1^2+C_2^2}$ of these terms is close to zero, we expect
$|\psi(T)\rangle =U(T,0)|0\rangle \simeq |0\rangle$. We find, as
shown in Fig.\ \ref{fig4}, that ${\mathcal N}$ comes very close to zero
(although it does not vanish, in contrast to exact numerics) around
$\omega_D \simeq 7.9$ which is remarkably close to the
freezing frequency observed in exact numerics.

\begin{figure}
{\includegraphics*[width=\linewidth]{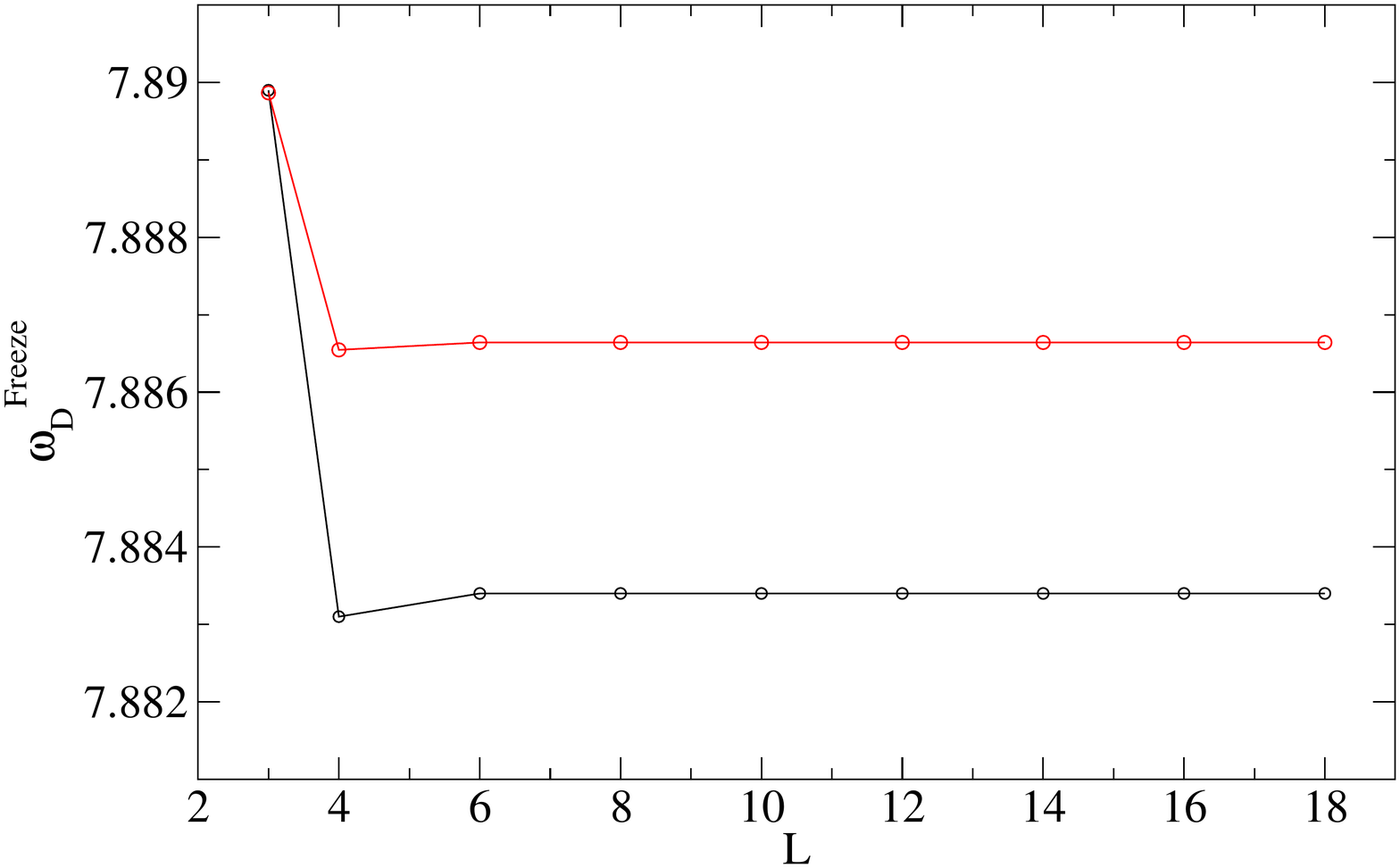}}
\caption{Plot of $\Delta_0/2$ (black solid line) and the highest freezing 
frequency $\omega_D^{\rm freeze}$ (red solid line) as a function of $L$. Here 
$\lambda=15$ and all units are the same as in Fig.\ \ref{fig1}.} \label{fig5} 
\end{figure}

The above perturbative analysis indicates that multiple spin terms
do not play an essential role in the freezing phenomenon since, at
least to $O(w^3)$, all of them annihilate the $|0\rangle$ state.
Thus it is natural to expect that this phenomenon can also be
qualitatively understood by focussing on small system sizes where
the small size of the Hilbert space allows for an exact analytical
calculation. To this end, we consider a $L=3$ system. In the $k=0$
sector, there are two states in its constrained Hilbert space. These
are $|0\rangle=|\downarrow \downarrow \downarrow \rangle$, and
$|1\rangle = (|\uparrow \downarrow \downarrow \rangle + |\downarrow
\uparrow \downarrow \rangle + |\downarrow
\downarrow \uparrow \rangle)/\sqrt{3}$. In the space of these states,
the Hamiltonian can be written, up to an irrelevant constant term, as
\begin{equation} H[\pm\lambda] =\left(\begin{array}{cc}
0 & -\sqrt{3}w\\
-\sqrt{3}w & \pm \lambda\\
\end{array}\right). \end{equation}
The Floquet Hamiltonian for this system can be computed exactly and is given by
\begin{eqnarray} H_{F}^{\rm exact}&=& \frac{ic}{T\sin (c)}\Big[ \frac{2\sqrt{3}
w \sin(\Delta_0 T/2)}{\Delta_0} \tau_x \nonumber\\
&& \quad \quad \quad \quad -\frac{4\sqrt{3}w\lambda\sin^2(\Delta_0 T/4)}{
\Delta_0^2} \tau_y \Big], \end{eqnarray}
where $\tau^{x,y}$ denotes Pauli matrices in the space of states
$|0\rangle$ and $|1\rangle$, $\Delta_0= \sqrt{12 w^2+\lambda^2}$ is
the static energy gap between the states $|0\rangle$ and $|1\rangle$
states, and $c=\cos^{-1} (1-24 w^2\sin^2(\Delta_0 T/4)/\Delta_0^2)$.
This leads to the expression for the matrix element between the
states $|0\rangle$ and $|1\rangle$ as
\begin{eqnarray} |\bra{0}H_F^{\rm exact} \ket{1}|= \frac{\omega_D}{2\pi}
\cos^{-1}\left[\frac{\lambda ^2+12 w^2 \cos \left(\Delta_0 T/2\right)}{
\Delta_0^2} \right]. \label{2state1} \end{eqnarray}
This shows that for $\omega_D= \omega_D^{\rm freeze}=\Delta_0/(2m)$,
where $m$ is an integer, the matrix element between $|0\rangle$ and
$|1\rangle$ exactly vanishes. Consequently, $|0\rangle$ does not
evolve at these frequencies. Thus the freezing frequencies are directly
related to the static gap $\Delta_0$ between the $|0\rangle$ and the
single up-spin ($|1\rangle$) states. We note that our analytic
expression for $\omega_D^{\rm freeze}$ provides a natural explanation
for the subharmonic structure of the lower freezing
frequencies. These frequencies turn out to match almost exactly with
ED based numerical computation for finite $L \le 26$. This is shown
in Fig.\ \ref{fig5} where $\Delta_0/2$ and the highest
$\omega_D^{\rm freeze}$ ($m=1$) is plotted as a function of $L$. The
reason for this near-perfect match is that multiple spin terms in
$H_F$ do not contribute to this phenomenon as explained earlier. We
also supplement this numerical check by an explicit analytic
calculation for $L=4$ in App.\ \ref{appB}.

\subsection{Sub-thermal steady state value}
\label{sub1}

In this section, we address the sub-thermal behavior of the system
as seen, for example, in the frequency range $ 5 \le
\omega_D \le 7.5$. Throughout this range, the dynamics
of the system, starting from the $|0\rangle$ state, is qualitatively
identical to that shown in Fig.\ \ref{fig2a} (d). It shows a rapid
approach to a steady state where $O_{22}$ assumes a sub-thermal
value; in addition, there are no persistent coherent oscillations, in
contrast to the $|\mathbb{Z}_2\rangle$ dynamics shown in Fig.\ \ref{fig2b} (d).
This behavior constitutes a novel route to a violation of ETH in
finite-sized systems for two reasons. First, we do not see here the
persistent oscillations usually seen in dynamics controlled by quantum scars,
and second, $O_{22}$ assumes sub-thermal, in contrast to
super-thermal, values in the steady state.

\begin{figure}
{\includegraphics*[width=\linewidth]{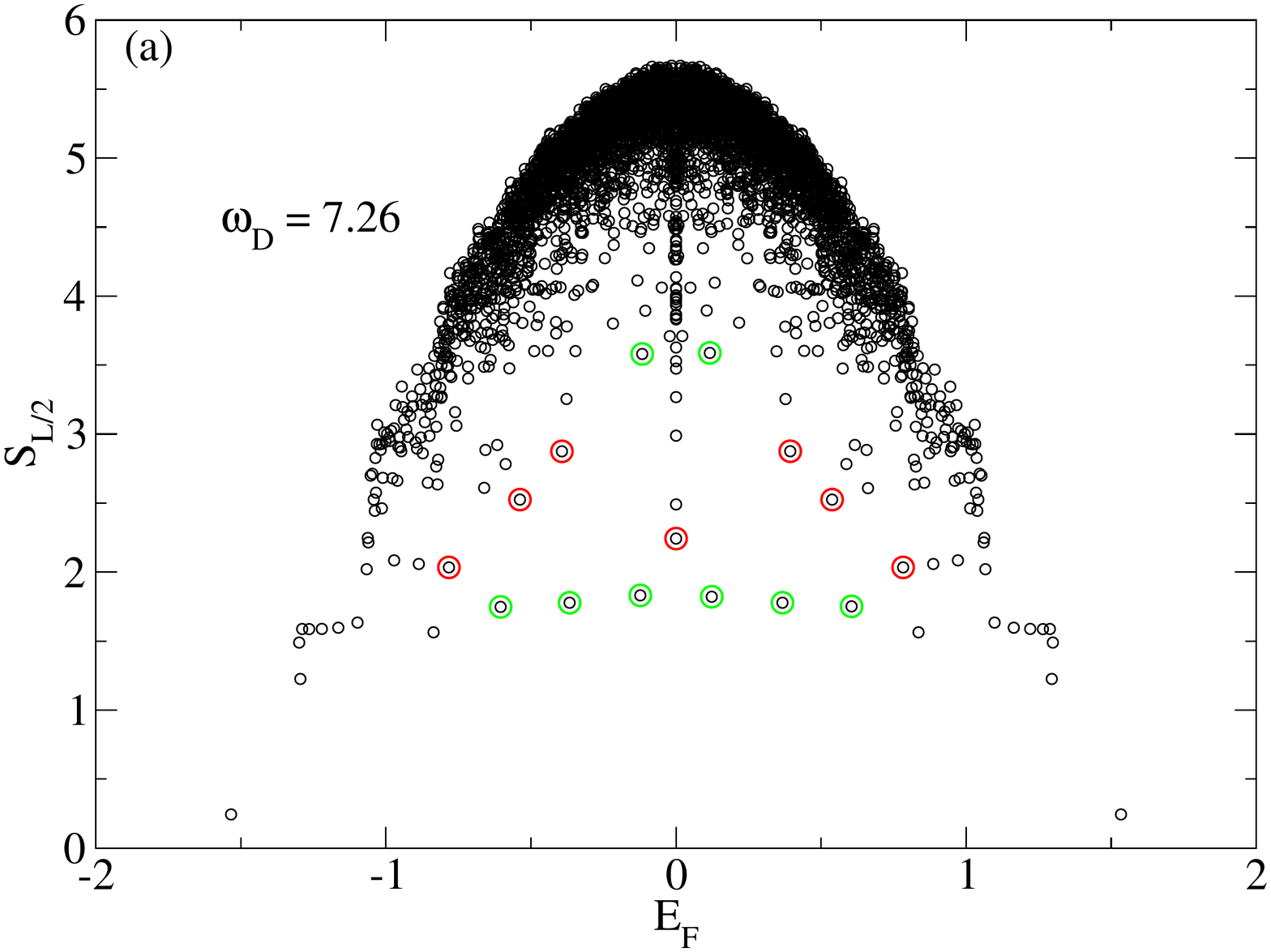}}\\
{\includegraphics*[width=\linewidth]{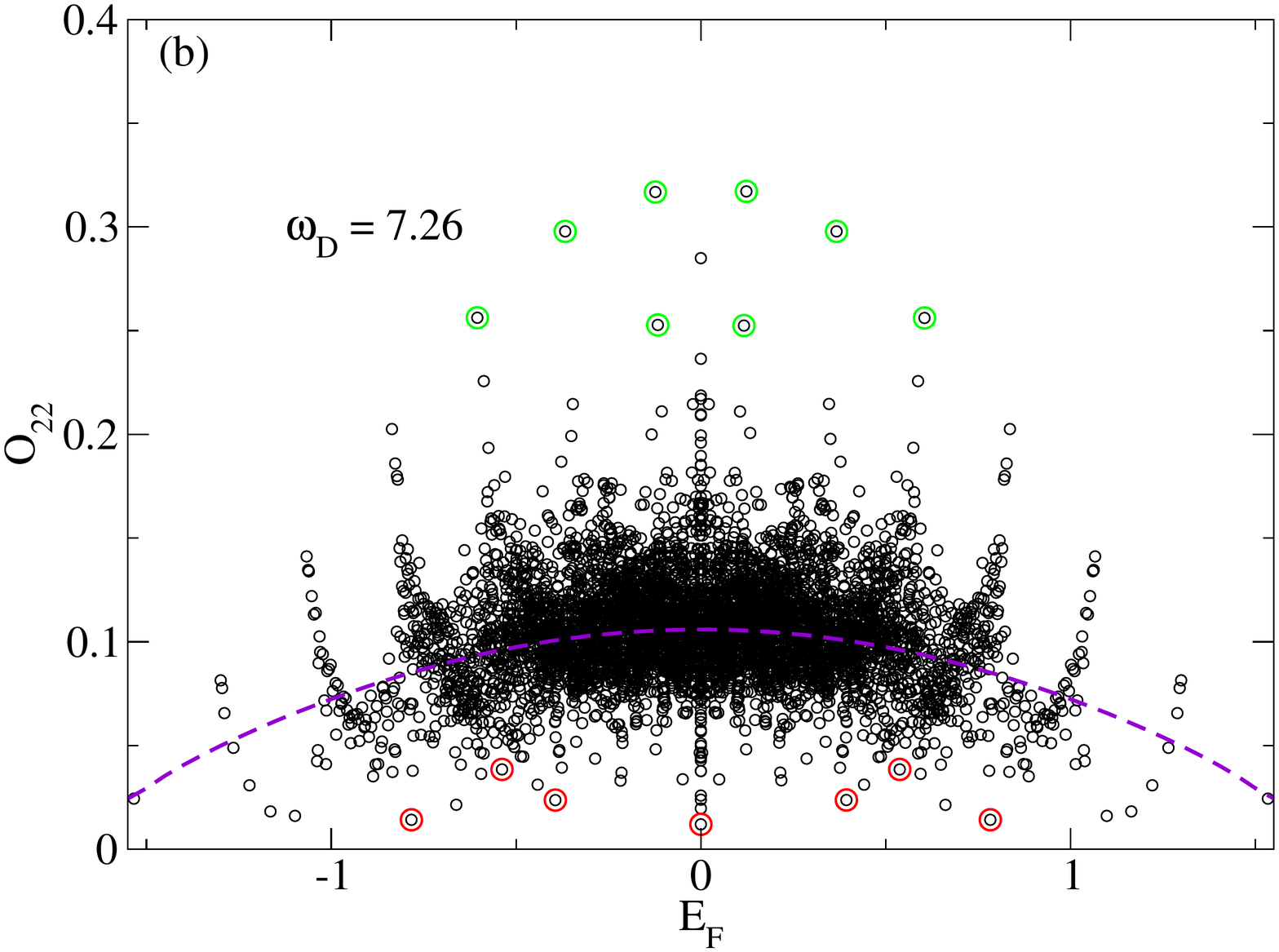}}
\caption{(a) Plot of $S_{L/2}$ for the eigenstates of $H_F$
for $L=26$ and $\omega_D =7.26$. The eigenstates with overlap
$>0.01$ with $|0\rangle ~(|Z_2\rangle)$ are shown using red (green) circles. 
(b) Plot for $O_{22}^m$ as a function of Floquet eigenstate quasienergies 
$\epsilon_F$. The violet dashed line indicates the ETH predicted value of 
$O_{22}$ at a temperature $T_0 (E_F)$. Here $L=26$, $\lambda=15$ and 
all units are the same as in Fig.\ \ref{fig1}.} \label{fig6} \end{figure}

To understand this phenomenon, we plot $S_{L/2}$ for the eigenstates
of $H_F$ for $\omega_D = 7.26$ in Fig.\ \ref{fig6} (a) and
$O_{22}^m$ as a function of Floquet quasienergies in Fig.\
\ref{fig6} (b). From both these panels, we note that there are
relatively few sub-thermal states with high overlaps with
the $|0\rangle$ state; in particular, there are no thermal
states with overlap $|\langle \chi_m|0\rangle|^2
>0.01$ at this frequency. Thus the weight of the
$|0\rangle$ state is distributed among a few sub-thermal and a
relatively large set of thermal states. The dynamics starting from
the $|0\rangle$ state within this frequency range is analogous to a
quantum system coupled to a bath which features a large range of
incommensurate natural frequencies; the presence of a large number
of thermal states which have small but finite overlaps with
$|0\rangle$ mimics the effect of a bath in the present context. The
presence of such a bath leads to fast decoherence of the oscillations
and leads to a steady state. This behavior is in stark contrast to
the dynamics at $\omega_D=8.5$ where a few sub-thermal states with
large overlaps control the dynamics.

\begin{figure}
{\includegraphics*[width=\linewidth]{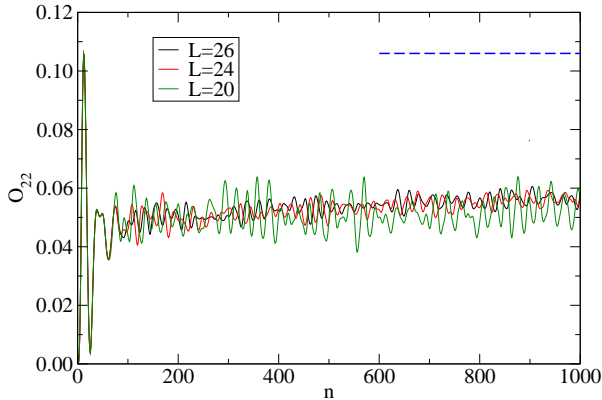}}
\caption{Plots of $O_{22}$ as a function of the number of drive cycles $n$ 
(stroboscopic time) starting from an initial state $|0\rangle$ for $L=20$ 
(green), $L=24$ (red), and $L=26$ (black), for $\omega_D=7.26$ and $\lambda
=15$. The blue dashed line indicates the ETH predicted thermal value of 
$O_{22}$. All units are the same as in Fig.\ \ref{fig1}.} \label{subthermal}
\end{figure}

The sub-thermal values of $O_{22}$ in this steady state are more
difficult to explain. Numerically, from Fig.\ \ref{fig6} (b), we
find that the Floquet eigenstates with relatively large overlaps with
the $|0\rangle$ state have sub-thermal values of $O_{22}^m$, and this
feature is opposite to that for states with large overlaps with
$|\mathbb{Z}_2\rangle$. From Eq.\ \eqref{decorr1}, we expect that
this feature will lead to sub-thermal values of $O_{22}$ in the
steady state. However, beyond this observation, we do not have a
more analytical explanation for this phenomenon. We also note that
such sub-thermal values of $O_{22}$ in the steady state are clearly
a finite-size effect; for $L \to \infty$, the number of thermal
Floquet eigenstates with finite overlap with $|0\rangle$ will be
exponentially larger than the sub-thermal states and their
contribution is expected to lead to the ETH predicted thermal value
of $O_{22}$ in the steady state. However, for all $L \le 26$ we do
not find thermal behavior; moreover for this range of system sizes,
the steady state value of $O_{22}$ remains almost constant as can be
seen from the values of $O_{22}^{DE}$ for $L=14, ~20,\,{\rm and}\,
24$ in Fig.\ \ref{fig1}. This indicates that a restoration of ETH is
expected only for $L \gg 26$; for $L\le 26$, we find a qualitatively
distinct and experimentally discernible characteristic of $O_{22}$
which is different from both scar-induced persistent oscillations
and ETH predicted thermalization. Furthermore, the behavior of
$O_{22}$ as a function of $n$ upto $1000$ drive cycles for different
system sizes (Fig.\ \ref{subthermal}) suggests that this non-ETH
sub-thermal behavior can persist as a prethermal regime for
reasonably large $n$ before the system eventually flows to an ITE
even for much larger system sizes. We therefore believe that this
phenomenon provides a novel route to ETH violation in a finite-sized
Rydberg chain.

\section{Discussion}
\label{diss}

In this work, we have studied the dynamics of a periodically driven
Rydberg chain starting from the $|0\rangle$ state using a square
pulse protocol. Our study involves exact numerics on finite-sized
chains and a perturbative Floquet Hamiltonian based analysis whose
analytic expression is derived using Floquet perturbation theory in
the high drive amplitude limit.

Our study indicate three distinct behaviors of such dynamics. First,
we show that dynamics starting from the $|0\rangle$ state can
exhibit scar-induced persistent oscillations over a range of drive
frequencies $8 \le \omega_D \le 12$. These scars are distinct from
their $\mathbb{Z}_2$ counterparts; they are absent in the high drive
frequency limit and are not eigenstates of the (renormalized) PXP
Hamiltonian studied earlier in the literature~\cite{scarfl1,scarref1,
scarref2a,scarref2b,scarref2c,scarref2d,scarref2e}. They coexist
with the $|\mathbb{Z}_2\rangle$ scars in the above-mentioned drive
frequency range. These $|0\rangle$ scars have higher entanglement
than their $|\mathbb{Z}_2\rangle$ counterparts and seem more fragile
to increasing system sizes. It will be worthwhile to understand
perturbations that may further reduce the entanglement of these
$|0\rangle$ scars.

Second, for specific drive frequencies $\omega_D^{\rm freeze} \simeq
7.88, 3.94, \cdots$, we find that the system exhibits dynamic freezing.
We provide an analytic, albeit qualitative, explanation of this
phenomenon using the perturbative Floquet Hamiltonian and supplement
it with exact analytic calculation at small system sizes $L=3$ and 4.
Our analysis relates the freezing frequencies with the energy gap
$\Delta_0$ between the $|0\rangle$ and $|1\rangle$ states:
$\omega_D^{\rm freeze}=\Delta_0/2m$ for $m=1,2,3,\cdots$. This
provides a natural explanation for the relation between several
freezing frequencies and also provides a reasonably accurate
estimate of these frequencies as can be seen by comparing the
analytic result with exact numerics for all $L \le 26$. We note that
such a behavior has no analogue for dynamics starting from the
$|\mathbb{Z}_2\rangle$ state. Such dynamical freezing should also be
visible in the thermodynamic limit as prethermal freezing which only
flows to an ITE after an exceptionally long time scale.

Third, we show that for $5 \le \omega_D \le 7.5$, the system reaches
a steady state with sub-thermal values of $O_{22}$ for all $L \le
26$ which constitutes a violation of ETH. In contrast to the
scar-induced weak violation of ETH, in this regime, the system does
not exhibit persistent oscillations for $O_{22}$. We relate this
behavior to the presence of small overlaps of a large number of
Floquet eigenstates with the $|0\rangle$ state; this leads to the
fast decay of coherent oscillations in $O_{22}$. Moreover,
numerically we find that $O_{22}^m$ for many of the Floquet
eigenstates which have high overlap with $|0\rangle$ assumes
sub-thermal values; this leads to sub-thermal values of $O_{22}$ in
the steady state. We note that such a violation of ETH is distinct
from its counterpart in the dynamics starting from the $|\mathbb{Z}_2\rangle$
state; it does not feature persistent oscillations
and leads to steady states with sub-thermal, rather than
super-thermal, values of $O_{22}$. Our numerical results for
finite-sized chains also suggest that this sub-thermal behavior can
survive as a prethermal phase for finite but large number of drive
cycles before the system eventually flows to an ITE in the
thermodynamic limit.

All the above three features should be observable in realistic
experiments with a Rydberg chain. The differences of our proposal with
experiments already carried out in Ref.\ \onlinecite{scarref1} are
two-fold. First, for our proposal, we need to start from the
$|0\rangle$ state. This is not difficult to implement since this
state turns out to be the ground state of $H$ in Eq.\ \eqref{ryd1} for
large positive $\Delta$. Second, we need to implement a periodic
variation of $\Delta$ according to the protocol given in Eq.\ \eqref{protocol1}
instead of a quench. Our prediction is that the dynamics starting from the
$|0\rangle$ state will show persistent scar-induced oscillations, dynamic
freezing and novel steady states featuring sub-thermal values of $C_{22} \equiv
O_{22}$ in such experiments.

In conclusion, we have shown that dynamics starting from the
$|0\rangle$ state in a finite-sized periodically driven Rydberg
chain shows scar-induced oscillations, dynamic freezing, and steady
states with sub-thermal value of correlators for various ranges of
drive frequencies which we have charted out. The first feature shows
that the Floquet Hamiltonian hosts two sets of coexisting scars; the
last two phenomena have no analogs in periodic or quench dynamics
involving the $|\mathbb{Z}_2\rangle$ initial state studied earlier. We have
provided an analytic, albeit perturbative, Floquet Hamiltonian which
explains these features qualitatively and have suggested experiments
which can test our theory.

On a broader level, our work suggests the possibility of interesting
prethermal Floquet phases at moderate and low drive frequencies in
the high drive amplitude limit~\cite{Vajna2018}. Such a
prethermalization mechanism is quite distinct from the well-known
long preheating times generated in Floquet systems when the driving
frequencies are much bigger than the local energy
scales~\cite{Saito_etal} and should lead to richer possibilities.

\begin{acknowledgments}
The work of A.S. is partly supported through
the Max Planck Partner Group program between the Indian Association
for the Cultivation of Science (Kolkata) and the Max Planck
Institute for the Physics of Complex Systems (Dresden). D.S. thanks
DST, India for Project No. SR/S2/JCB-44/2010 for financial support.
\end{acknowledgments}

\appendix

\section{Floquet perturbation theory}
\label{appA}

In this appendix, we sketch the essential points regarding
the derivation of the Floquet Hamiltonian. For this purpose, we first
note that, as explained in the text, $U_0$ is given by
\begin{eqnarray} U_0 (t,0) &=& e^{i \lambda t \sum_j \sigma_j^z /2} \quad
{\rm for} \quad t\le T/2, \nonumber\\
&=& e^{i \lambda (T-t) \sum_j \sigma_z^j/2} \quad {\rm for} \quad
T/2 \le t \le T. \label{u0eqapp} \end{eqnarray}
Noting the form of $U_0$, we define the functions
\begin{eqnarray} I_0^{(1)} (t,s) &=& e^{i\lambda st}, \quad I_0^{(2)} (t,T,s)
= e^{i\lambda s(T-t)}, \nonumber\\
I_1(t,s) &=& \int_0^t dt I_0^{(1)} (t,s) = \frac{1}{i\lambda s}
\left( e^{i\lambda s t}-1\right), \label{fn1app} \\
I_2(t,T,s) &=& \int_0^t dt I_0^{(2)} (t,T,s) = \frac{e^{i\lambda s
T}}{i\lambda s} \left(1- e^{-i\lambda s t}\right). \nonumber
\end{eqnarray}
where $s=\pm 1$. The first order term in the Floquet Hamiltonian
$U_1 = -i \int_0^T U_0^{\dagger} (t,0) H_1 U_0(t,0) dt$ can be
expressed in terms of these functions as
\begin{eqnarray} \langle U_1(T,0)\rangle_{m,n} &=& i w \sum_{s=\pm}
\delta_{m,n+s} \Big[\int_0^{T/2} dt I_0^{(1)} (t,s) \nonumber\\
&& + \int_{T/2}^T dt I_0^{(2)} (t,T,s) \Big] \nonumber\\
&=& i w \delta_{m,n+s} \left[ I_1(T/2,s) + I_2(T,T,s) \right] \nonumber\\
&=& \frac{2w}{\lambda s} \delta_{m,n+s} \left( e^{i\lambda s T/2}-1\right),
\label{u1matelapp} \end{eqnarray}
which leads to the ${\rm O} (w)$ term in the Floquet Hamiltonian
(Eq.\ \eqref{fl1}) as discussed in the main text.

Next, we evaluate the matrix elements of $U_2(T,0) = (-i)^2 \int_0^T
dt_1 H_I(t_1) \int_0^{t_1} dt_2 H_I(t_2)$, where $H_I= U_0^{\dagger}
H_1 U_0$ is the perturbative term in $H$ in the interaction picture,
between two states $m$ and $n$. A straightforward calculation shows
\begin{widetext}
\begin{eqnarray} \langle U_2(T,0)\rangle_{mn} &=& (iw)^2 \sum_{s_1=\pm,s_2=\pm}
\delta_{m,n+s_1+s_2} \left[ \left( \int_0^{T/2} dt I_0^{(1)} (t_1,s_1)
+ \int_{T/2}^T dt I_0^{(2)} (t_1,T,s_1) \right) \right.\nonumber\\
&& \times \left. \left( \int_{0}^{t_1} dt_2 I_0^{(1)} (t_2,s_2)
\theta(t_1-T/2) + \int_{T/2}^{t_1} dt_2 I_0^{(2)} (t_2,T,s_2)
\theta(T/2-t_1)\right) \right]. \label{secordintapp} \end{eqnarray}
\end{widetext}
To evaluate this expression we first define the integrals
\begin{widetext}
\begin{eqnarray}
I_3(t,s_1,s_2) &=& \int_0^{t} dt_1 I_0^{(1)} (t_1,s_1) I_1 (t_1,s_2)
= - \frac{1}{2\lambda^2} \left[ \left(e^{i\lambda s_1 t}-1\right)^2
\delta_{s_1,s_2} + \left(1-e^{i \lambda s_1 t} + i \lambda s_1 t
\right) \delta_{s_1,-s_2} \right], \label{i3i4app} \\
I_4(t,T,s_1,s_2) &=& \int_{T/2}^{t} dt_1 I_0^{(2)} (t_1,T,s_1) I_2
(t_1,T,s_2) = - \frac{1}{2\lambda^2} \left[ e^{i\lambda s_1 (2t-T)}
\left(e^{i\lambda s_1 t}- e^{i\lambda s_1 T/2}\right)^2
\delta_{s_1,s_2} \right. \nonumber\\
&& \quad \quad \quad \quad \quad \quad \quad \quad \quad \quad \quad \quad
~\quad \quad \quad \quad \quad \quad \left.+ \left(2(1-e^{i
\lambda s_1 (2t-T)}) - i \lambda s_1 (2t-T) \right) \delta_{s_1,-s_2} \right].
\nonumber \end{eqnarray}
\end{widetext}
In terms of these integrals, we can write
\begin{eqnarray} U_2(T,0) &=& \sum_{i,j} \sum_{s_1,s_2=\pm} c^{(2)}_{s_1 s_2}
\sigma_i^{s_1} \sigma_j^{s_2}, \nonumber\\
c_{s_1 s_2}^{(2)} &=& -w^2 \left[ I_3(T/2,s_1,s_2) ~+~
I_4(T,T,s_1,s_2) \right. \nonumber\\
&& \left. ~~~~~~+~ I_2(T,T,s_1) I_1(T,s_2) \right]. \label{order2coeffapp}
\end{eqnarray}
Using Eqs.\ \eqref{order2coeffapp} and \eqref{i3i4app}, we find that
\begin{eqnarray}
c_{s,s}^{(2)} &=& \frac{2w^2}{\lambda^2} \left(e^{i \lambda s
T/2}-1\right)^2, \nonumber\\
c_{s,-s}^{(2)} &=& -\frac{8w^2}{\lambda^2} \sin^2 (\lambda T/4),
\label{c2app} \end{eqnarray}
which are used in the main text.

Finally, we address the third order term given in Eq.\ \eqref{thirdorderu}.
In terms of the functions $I_0^{(1)}$ and $I_0^{(2)}$ these can be written as
\begin{widetext}
\begin{eqnarray} \langle U_2(T,0)\rangle_{mn} &=& (-iw)^3
\sum_{s_1,s_2,s_3=\pm} \delta_{m,n+s_1+s_2+s_3} \left[
\left( \int_0^{T/2} dt I_0^{(1)} (t_1,s_1) + \int_{T/2}^T dt
I_0^{(2)} (t_1,T,s_1) \right) \right.\nonumber\\
&& \times \left. \left( \int_{0}^{t_1} dt_2 I_0^{(1)} (t_2,s_2)
\theta(t_1-T/2) + \int_{T/2}^{t_1} dt_2 I_0^{(2)} (t_2,T,s_2)
\theta(T/2-t_1)\right) \right. \nonumber\\
&& \left. \times \left( \int_{0}^{t_2} dt_3 I_0^{(1)} (t_3,s_3)
\theta(t_2-T/2) + \int_{T/2}^{t_2} dt_3 I_0^{(2)} (t_3,T,s_3)
\theta(T/2-t_2)\right)\right]. \label{thirdintapp} \end{eqnarray}
\end{widetext}
Evaluating Eq.\ \eqref{thirdintapp} carefully after taking care of all
commutators, we obtain
\begin{eqnarray} U_{3} (T,0) &=& \sum_{j,j',j"} \sum_m \sum_{s_1,s_2,s_3=\pm}
c^{(3)}_{s_1 s_2 s_3} \tilde \sigma_{j}^{s_1} \tilde
\sigma_{j'}^{s_2} \tilde \sigma_{j"}^{s_3}, \nonumber\\
c^{(3)}_{s_1,s_2,s_3} &=& \sum_i \alpha_i(T,s_1,s_2,s_3), \nonumber\\
\alpha_1(T,s_1,s_2,s_3) &=& \int_0^{T/2} dt I_0^{(1)} (t_1,s_1)
I_3(t,s_1,s_2), \nonumber\\
\alpha_2(T,s_1,s_2,s_3) &=& \int_{T/2}^T dt I_0^{(2)} (t_1,Ts_1)
I_4(t,T,s_1,s_2), \nonumber\\
\alpha_3(T,s_1,s_2,s_3) &=& I_2(T,T,s_1) I_3(T/2,s_2,s_3), \label{order3int} \\
\alpha_4(T,s_1,s_2,s_3) &=& I_2(T,T,s_1) I_2(T,T,s_2) I_1(T/2,s_3).
\nonumber \end{eqnarray}
The integrals $\alpha_{1,2}$ lead to cumbersome expressions but can
be straightforwardly obtained using Mathematica. Evaluating these
integrals and using Eqs.\ \eqref{fn1app} and \eqref{i3i4app}, we finally
obtain the coefficients $c_{s_1,s_2,s_3}^{(3)}$ used in the main text.

\section{Freezing frequencies for $L=4$}
\label{appB}

In this Appendix, we consider a $L=4$ system and provide an analytical
expression for the freezing frequencies. It turns out that $L=4$ is the lowest
value of $L$ for which the Hilbert space has a two-particle manifold and has
dimension $6$. The $k=0$ sector has only three states given by
\begin{eqnarray} \ket{0}&=&\ket{0000}, \nonumber\\
\ket{1}&=&\frac{1}{2} (\ket{1000}+\ket{0100}+\ket{0010}+\ket{0001}), \nonumber\\
\ket{2}&=&\frac{1}{\sqrt{2}} (\ket{1010}+\ket{0101}),
\label{eq1} \end{eqnarray}
where $|0\rangle ~(|1\rangle)= |\downarrow\rangle ~(|\uparrow \rangle)$. In
this subspace the Hamiltonian $H_{\rm spin}[\pm \lambda]$ can be written as
\begin{equation} H_{\pm}=\left(\begin{array}{ccc}
0 & 2w & 0\\
2w & \pm \lambda & \sqrt{2}w\\
0 & \sqrt{2}w & \pm 2\lambda\\
\end{array}\right). \label{eq2} \end{equation}
It is quite difficult to explicitly calculate $H_F$ from Eq.\
\eqref{eq2} for the square pulse protocol. Instead we will provide an analytic
expression for $U_{11} (T,0)$. It is easy to see that this is equivalent to
computing $|\bra{0}H_F\ket{1}|$ for the purpose of locating the freezing
frequencies. The argument goes as follows. The fidelity after one cycle
$|\bra{0}U(T)\ket{0}|^2=U_{11}$ should be unity at the freezing points.
So $U_{11} (T,0)=1$ acts as an identifier of the freezing points.

To this end, we first compute the eigenspectrum of $H[\pm \lambda]$.
The eigenvalues of $H[\lambda]$ are given by
\begin{eqnarray} \epsilon_1 &=& w ~[n ~+~ 2R^{1/3}\cos(\alpha/3)], \nonumber\\
\epsilon_2 &=& w ~[n ~-~ 2R^{1/3}\cos((\alpha+\pi)/3)], \nonumber\\
\epsilon_3 &=& w ~[n ~-~ 2R^{1/3}\cos((\alpha-\pi)/3)], \label{eigen4}
\end{eqnarray}
where $n=\lambda/w$ , $R = (n^2 /3 + 2)^{3/2}$, and $\alpha =
\tan^{-1} (-R/n)$. It turns out that the eigenvalues of $H[-\lambda]$ are
$-\epsilon_{1,2,3}$. The corresponding normalized eigenstates of
$H[\pm \lambda]$ are given by
\begin{eqnarray} \ket{\psi^{\pm \lambda}_i}&=& \frac{1}{N_i}
\left(\begin{array}{c} 1\\ \pm \epsilon_i/2w\\
(1-\frac{2\lambda}{\epsilon_i})^{-1}/\sqrt{2} \end{array}\right), \nonumber\\
N_i &=& \sqrt{1+\frac{\epsilon_i^2}{4w^2}+\frac{\epsilon_i^2}{2(\epsilon_i
-2\lambda)^2}}. \end{eqnarray}

\begin{figure}
\includegraphics[width=\linewidth]{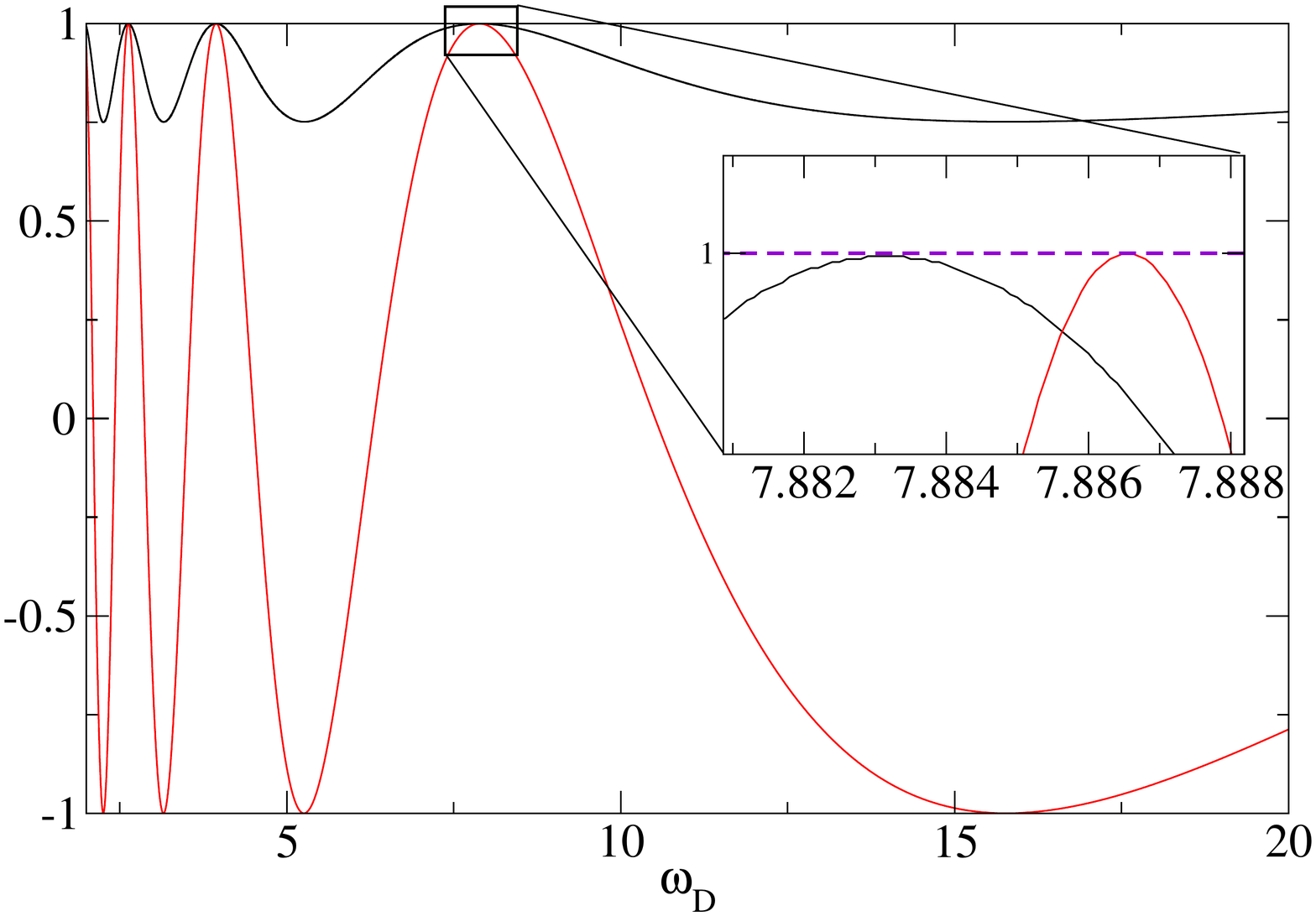}
\caption{Curves for $U_{11} (T,0)$ using Eq.\ \eqref{U11eq} (in black) and 
$\cos(\Delta_0 T/2)$ (in red) as a function of $\omega_D$ show that the drive 
frequencies at which $U_{11} (T,0)=1$ and $\cos(\Delta_0T/2)=1$ almost coincide
(see inset for details) when $\lambda \gg w$.} \label{L4} \end{figure}

Using the eigenspectrum of $H[\pm \lambda]$, we can obtain an
expression for $U_{11} (T,0)$. A straightforward calculation yields
\begin{widetext}
\begin{eqnarray}
U_{11}&=&\frac{1}{4}\Big(\frac{\epsilon_1^2 \left(\frac{2}{(\epsilon_1-2
\lambda )^2}-\frac{1}{w^2}\right)+4}{N_1^4}+\frac{\epsilon_2^2
\left(\frac{2}{(\epsilon_2-2 \lambda)^2}-\frac{1}{w^2}\right)+4}{N_2^4}+
\frac{\epsilon_3^2 \left(\frac{2}{(\epsilon_3-2 \lambda )^2}-\frac{1}{w^2}
\right)+4}{N_3^4}\nonumber\\
&+&\frac{2 \cos \left(\frac{1}{2} T (\epsilon_1-\epsilon_3)\right)
\left(\epsilon_1 \epsilon_3 \left(\frac{2}{(\epsilon_1-2 \lambda ) (\epsilon_3
-2 \lambda)}-\frac{1}{w^2}\right)+4\right)}{N_1^2 N_3^2}+\frac{2 \cos
\left(\frac{1}{2} T (\epsilon_2-\epsilon_3) \right) \left(\epsilon_2 \epsilon_3
\left(\frac{2}{(\epsilon_2-2 \lambda ) (\epsilon_3-2 \lambda )}-\frac{1}{w^2}
\right)+4\right)} {N_2^2 N_3^2}\nonumber\\
&+&\frac{2 \cos \left(\frac{1}{2} T (\epsilon_1-\epsilon_2)\right)
\left(\epsilon_1 \epsilon_2 \left(\frac{2}{(\epsilon_1-2 \lambda ) (\epsilon_2
-2 \lambda )}-\frac{1}{w^2}\right)+4\right)}{N_1^2 N_2^2}\Big). \nonumber\\
\label{U11eq}
\end{eqnarray}
\end{widetext}
Analyzing this expression, we find that $U_{11} (T,0) \simeq 1$ for 
$\cos \left( \Delta_0 T/2 \right)=1$, where $\Delta_0=\epsilon_2-\epsilon_1$ 
in spite of the complicated expression for $U_{11} (T,0)$ (see 
Fig.\ \ref{L4}) when $\lambda \gg w$. This yields
$\omega_D^{\rm freeze}= \Delta_0/(2m)$. Thus the freezing condition is
identical to the result for $L=3$ given in the main text.

\section{FSA from $|0\rangle$}
\label{appC}

In this appendix, we shall chart out the FSA formalism using $|0\rangle$ as 
the starting state. In Sec.\ \ref{fsa1}, we demonstrate the computation using 
the PXP model while in Sec.\ \ref{fsa2}, we add higher spin terms to the 
Hamiltonian used for the FSA analysis.

\subsection{FSA in the PXP model}
\label{fsa1}

\begin{figure}
\includegraphics[width=\linewidth]{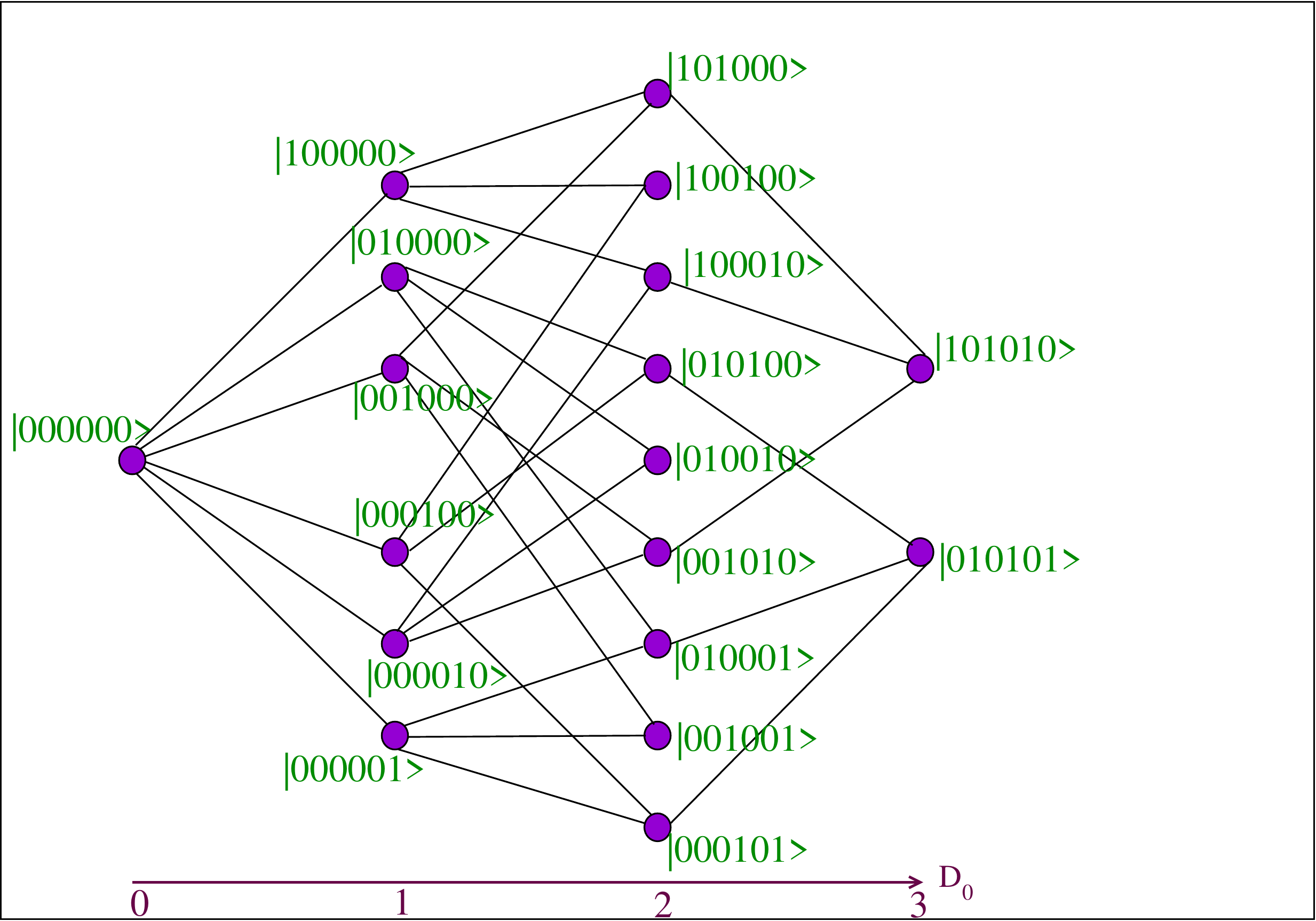}
\caption{Graph representation of the action of PXP model in the constrained 
Hilbert space for $L=6$. The Hamming distance ($D_0$) of any configuration is 
measured from the state $|0\rangle$. Here $1 ~(0) \equiv \uparrow 
(\downarrow)$.} \label{FSAgraph} \end{figure}

In this section we carry out the FSA analysis using the PXP model
starting from $|0\rangle$. We decompose the PXP Hamiltonian
\begin{equation} H_{PXP}=w\sum_{i=1}^L\tilde{\sigma}^x_i=w\sum_i(
\tilde{\sigma}^{+}_i+\tilde{\sigma}^{-}_i)=H^{+}+H^{-}. \end{equation}
We assume $w=1$ and hence drop it from all further calculations in
this section. This decomposition is motivated by the requirement
that $H^{-}$ should annihilate the initial state ($|0\rangle$) and
$H^{+}$ should annihilate the final state ($|{\mathbb Z}_2\rangle$
and $|\bar{{\mathbb Z}_2}\rangle$). This leads to the modified
Lanczos operation acting in the Krylov space spanned by $L/2+1$
states (for system size $L$) obtained via repeated action of $H^{+}$
on the state $|0\rangle$, as is customary in the FSA method. One can
reach the final states by acting $H^{+}$ $L/2$ times on the initial
state (see Fig.\ \ref{FSAgraph}); these operations define the number
of steps in the FSA method. The dynamics can then be visualized as a
coherent forward and backward scattering in between these two
extreme states.

We start from $|v_0\rangle=|0\rangle$. We note that $H_{-}|v_0\rangle=0$ 
whereas
\begin{eqnarray} H^{+}|0\rangle &=& \sum^L_{i=1}\tilde{\sigma}^{+}_i|0\rangle =
\sum^L_{i=1} |\cdots\uparrow_i\cdots\rangle = \sum^L_{i=1} |i\rangle ,
\end{eqnarray}
where the ellipsis indicate down spins, and here and in the rest of
this Appendix, we shall use the site index (indices) of the up-spin
(up-spins) to label a state. In this analysis, we use periodic
boundary condition so that the site index is always a number modulo
$L$, and $\beta_1=\sqrt{|| H^{+}|0\rangle ||} =\sqrt{L}$. This leads to
\begin{equation} |v_1\rangle=\frac{1}{\beta_1}
H^{+}|0\rangle=\frac{1}{\sqrt{L}}\sum^L_{i=1} |i\rangle . \end{equation}
One can easily check that $H^{-}|v_1\rangle=\beta_1|v_0\rangle$. Thus
the action of $H^{+}$ can be totally undone by $H^{-}$ which means
that the first FSA step is exact; no error is introduced at this step.

Next, we act $H^{+}$ on $|v_1\rangle$ below
\begin{eqnarray} H^{+}|v_1\rangle &=& \frac{1}{\sqrt{L}}\sum^L_{i=1}
\tilde{\sigma}^+_i \sum^L_{j=1} |j\rangle\nonumber\\
 &=&\frac{1}{\sqrt{L}}\sum_{i\neq (j-1),j,(j+1)} \ \sum^L_{j=1}|i,j\rangle .
\label{Hplusv1} \end{eqnarray}
The state in Eq.\ \eqref{Hplusv1} is a superposition of $L(L-3)/2$
distinct two up-spin states. This allows us to write
\begin{eqnarray} \beta_2 &=& \sqrt{||H^{+}|v_1\rangle||} = \sqrt{2(L-3)}.
\label{betaeq1} \end{eqnarray}
and leads to
\begin{eqnarray} |v_2\rangle&=&\frac{1}{\beta_2}H^{+}|v_1\rangle = 
\sqrt{\frac{2}{L(L-3)}} \sum_{ij} |i,j\rangle , \end{eqnarray}
where the sum is over all distinct two up-spin states in the
constrained Hilbert space. Again, we can easily check
$H^{-}|v_2\rangle=\beta_2|v_1\rangle$ which means the FSA is exact in
the second step.

Next, we act with $H^{+}$ on $|v_2\rangle$ which gives
\begin{eqnarray} &&H^{+}|v_2\rangle = \sqrt{\frac{2}{L(L-3)}} \Big(\sum_{i\neq
(j-1),j,j+1,j+3} \sum_{j=1}^L|i,j,j+2\rangle \nonumber\\
&& + \sum_{i\neq j,j-1,j+2, j'-1,j',j'+1} \ \sum_{|j'-j|\geq 3}
\sum_{j=1}^L |i,j,j'\rangle \Big). \label{Hplusv2} \end{eqnarray}
A careful counting of the number of states in Eq.\ \eqref{Hplusv2} leads to 
$L(L-4)(L-5)/2$ three-up spin distinct states in Eq.\ \eqref{Hplusv2} with
equal participation. But the total number of distinct three up-spin states
in the constrained Hilbert space is $L(L-4)(L-5)/6$. Thus we find
\begin{eqnarray} \beta_3=\sqrt{||H^{+}|v_2\rangle||}= \sqrt{\frac{3(L-4)
(L-5)}{L-3}}, \end{eqnarray}
so that
\begin{eqnarray} |v_3\rangle &=& \frac{H^{+}|v_2\rangle}{\beta_3} \nonumber\\
&=& \sqrt{\frac{6}{L(L-4)(L-5)}} \sum_{j_1,j_2,j_3} |j_1,j_2,j_3\rangle
\end{eqnarray}
where the sum is taken over all distinct three up-spin states within
the constrained Hilbert space. This yields
\begin{equation} H^{-}|v_3\rangle=\sqrt{\frac{6}{L(L-4)(L-5)}}\sum_{j_1,j_2}
|j_1,j_2\rangle . \label{Hminusv3} \end{equation}
The number of states with two up-spins in Eq.\ \eqref{Hminusv3} is
$L(L-4)(L-5)/2$ which is different from the number of two up-spin
states obtained in the second FSA step ($L(L-3)/2$). The ratio of
the number of these states is $(L-4)(L-5)/(L-3)$ which is not an
integer for all values of $L$. Thus $H^{-}|v_3\rangle$ is not
proportional to $|v_2\rangle$ and the action of $H^{+}$ on $|v_2\rangle$ 
cannot be completely undone by $H^{-}$, i.e., $H^{-}|v_3\rangle\neq \beta_3 
|v_2\rangle$.

Thus we see that the third FSA step introduces errors; such errors
are introduced in all $n \ge 3$ steps. We define the error, $\delta_n$,
introduced in the $n^{\rm th}$ step as
\begin{equation} \delta_n=||H^{-}|v_n\rangle-\beta_n|v_{n-1}\rangle||.
\label{error} \end{equation}
In this notation $\delta_1=\delta_2=0$ and $\delta_n \neq 0$ for
$n\geq 3$. We note that in the PXP model there is no parameter to
tune which may minimize such errors. This indicates the instability
of this procedure which becomes more apparent with increasing $L$
where there are more FSA steps. This indicates our need to go beyond
the PXP model to find stable $|0\rangle$ scars.

In the next section we show that the inclusion of non-trivial three-spin 
terms such as the one obtained in ${\rm O} (w^3)$ perturbation theory in 
Sec.\ \ref{floquet} of the main text provides a way to minimize such errors.

\subsection{FSA in the modified PXP model}
\label{fsa2}

In this section we reformulate the FSA by adding a three-spin term
similar to that found using Floquet perturbation theory (but with an
arbitrary real coefficient $h$) for the bare PXP model. The total
Hamiltonian is now
\begin{eqnarray} H_1(h) \equiv H_1 &=& \sum_{i=1}^L\tilde{\sigma}^x_i+h
\sum_{i=1}^L(\tilde{\sigma}^+_i\tilde{\sigma}^-_{i-1}\tilde{\sigma}^-_{i+1}
+{\rm H.c.}). \nonumber\\
\end{eqnarray}
This can be decomposed in the similar manner as before into
$H^-_{1}$ and $H^+_{1}$, where
\begin{equation} H^-_{1}=\sum_{i=1}^L\tilde{\sigma}^-_i+h\sum_{i=1}^L
\tilde{\sigma}^+_i\tilde{\sigma}^-_{i-1}\tilde{\sigma}^-_{i+1}, \end{equation}
and $H^+_{1}=(H^-_{1})^\dagger$. We start again with
$|v_0\rangle=|0\rangle$. As the additional term in $H_1$ annihilates
$|0\rangle$, the first FSA step remain unchanged, i.e.,
$\beta^{n}_1=\sqrt{L}$ and $|v^{n}_1\rangle=\sum_i|i\rangle/\sqrt{L}$.

The second FSA step is more complicated. Here we have
\begin{widetext}
\begin{eqnarray} H^+_{1}|v^{n}_1\rangle&=&\frac{2}{\sqrt{L}} 
\sum_{i\ne j,j-1,j+1} \sum_j |i,j\rangle ~+~ \frac{h}{\sqrt{L}} \sum_{i=1}^L
|i-1,i+1\rangle \nonumber\\
&=&\frac{1}{\sqrt{L}}\left[(2+h) \sum_{i=1}^L|i-1,i+1 \rangle ~+~ 2
\sum_{j\neq(i-2),(i-1),i,(i+1),(i+2)} \ \sum_{i=1}^L|i,j\rangle \right].
\label{Hplusnewvnew1} \end{eqnarray}
\end{widetext}
In Eq.\ \eqref{Hplusnewvnew1}, the second term on the right
represents states with two up-spins that are separated by at least
two lattice sites. The norm $\beta^{n}_2$ of this state is given by
\begin{eqnarray} \beta^{n}_2&=&\sqrt{||H^+_{1}|v^{n}_1\rangle||} =
\sqrt{h^2+4h+(2L-6)}. \nonumber\\
\end{eqnarray}
This allows us to obtain the new FSA vector in the second step,
$|v^{n}_2\rangle= H^+_{1}|v^{n}_1\rangle /\beta_2^{n}$. It is
possible to check that
$H^-_{1}|v^{n}_2\rangle=\beta_2^{n}|v_1^{n}\rangle$. This shows that
the FSA is error-free up to the second step although the norm and the
FSA vectors are modified due to the presence of the three-spin term in $H_1$.

We now show that the third FSA step leads to the first non-trivial
error. Calculating the action of $H^+_{1}$ on $|v_2^{n}\rangle$ is
straightforward but cumbersome. After grouping all the similar classes of
states generated by the action of different terms in $H^+_{1}$ and
summing the corresponding coefficients, we get
\begin{widetext}
\begin{eqnarray} H^+_{1}|v^{n}_2\rangle &=&\frac{6}{\sqrt{L}\beta^{n}_2} 
\left[\sum_{i=1}^L (h+1)|i,i+2,i-2\rangle+
(\frac{h}{2}+1) \sum_{j\neq (i-2)..(i+4)} \ \sum_{i=1}^L |i,i+2,j\rangle +
\sum_{|i-j|\geq 2, |j-k| \geq 2, |i-k|\geq 2} |i,j,k\rangle \right].
\nonumber\\
\label{Hplusnewvnew2} \end{eqnarray}
\end{widetext}
The last summation in Eq.\ \eqref{Hplusnewvnew2} is over those three
up-spin states that have no two up-spins as nearest neighbors. The
number of such states is $L(L-7)(L-8)/6$. A straightforward
calculation, similar to those presented earlier, yields
\begin{eqnarray} \beta^{n}_3 &=&\sqrt{||H^+_{1}|v^{n}_2\rangle||} \\
&=& \frac{\sqrt{9(L-3)h^2+36(L-5)h+6(L-4)(L-5)}}{\beta^{n}_2}. \nonumber
\end{eqnarray}
Thus we find that the third FSA vector $|v^{n}_3\rangle=
H^+_{1}|v^{n}_2\rangle/\beta_3^n$.

\begin{figure}
\includegraphics[width=\linewidth]{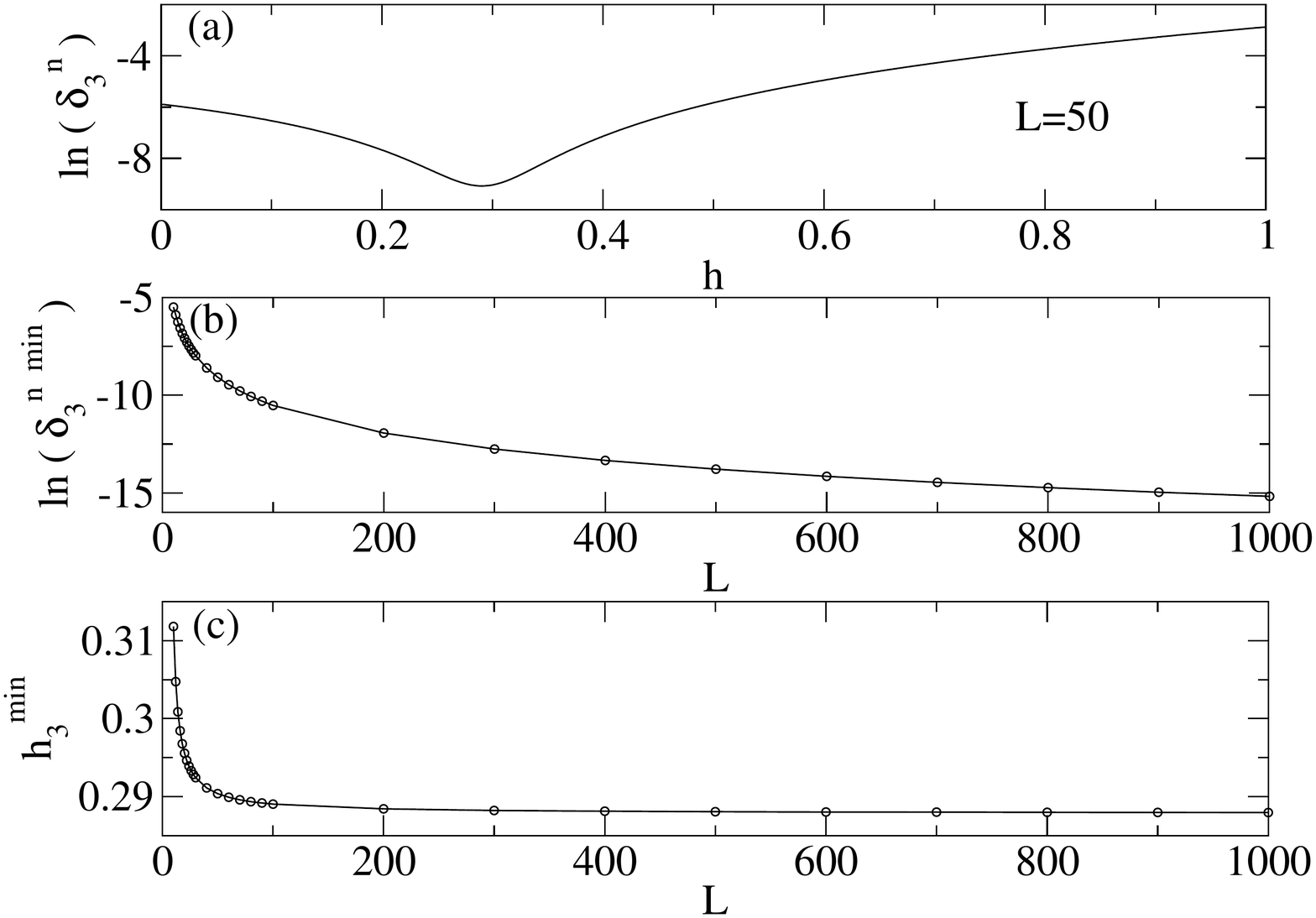}
\caption{(a) Plot of $\ln(\delta_3^n)$ as a function of $h$ for
$L=50$ showing a minima at $h_3^{{\rm min}} \simeq 0.3$. (b) Plot of
$\ln(\delta^{n\, {\rm min}}_3)$ as a function of $L$. (c) Plot of
$h_3^{{\rm min}}$ as a function of $L$ showing its saturation at
large $L$.} \label{3rdFSAplot} \end{figure}

Next we will calculate $H^-_{1}|v^{n}_3\rangle$. This is
straightforward but again involves a complicated counting of states.
Here we present the final expression,
\begin{widetext}
\begin{eqnarray} H^-_{1}|v^{n}_3\rangle &=&\frac{1}{\sqrt{L}\beta^{n}_2
\beta^{n}_3} \Big[ (12(h+1)+3(L-7)(h+2)) \sum_{i=1}^L|i,i+2\rangle
+(12h^2+18h+6L-36)\sum_{i=1}^L|i,i+3\rangle \nonumber\\
&& ~~~~~~~~~~~~~~+ (6h^2+24h+6L-36) \sum_{j\neq(i-3)..(i+3)} \sum_{i=1}^L
|i,j\rangle \Big]. \end{eqnarray}
\end{widetext}
It is easy to see that $H^-_{1}|v^{n}_3\rangle \neq \beta^{n}_3|v^{n}_2
\rangle$. The following norm quantifies the error
\begin{widetext}
\begin{eqnarray} \delta^n_3&=&||H^-_{1}|v^{n}_3\rangle-\beta^{n}_3|v^{n}_2
\rangle|| = \frac{f(h,L)}{g(h,L)}, \nonumber\\
f(h,L)&=&\big[(6L+6)h^6+(72L-168)h^5+(300L-1368)h^4+(288L-1608)h^3 \nonumber\\
&& +(30L-438)h^2+(-120L+696)h+(24L-120)\big], \label{fsaerrorn} \\
g(h,L)&=& \big[(3L-9)h^4+(24L-96)h^3+(8L^2-6L-146)h^2+(32L^2-264L+520)h
+4L^3-48L^2+188L-240 \big]. \nonumber
\end{eqnarray}
\end{widetext}
It is easy to see that $\lim_{L \to \infty} \delta^n_3(h,L) \to 0 $
for all $h$ which means that in the thermodynamic limit, the error
vanishes. But in a finite system, this error must be minimized to
enhance the oscillation from $|0\rangle$. The result of such a
minimization is shown in Fig.\ \ref{3rdFSAplot} (a) where
$\delta_3^n$ is plotted as a function of $h$ for $L=50$; we find that
it indeed shows a minima at a non-zero $h=h_3^{\rm min}$. This points
out the importance of the three-spin term in the Hamiltonian; its
coefficients provide us with the necessary control knob for
minimization of the FSA error leading to maximization of scar-induced
oscillations. We note that such a term does not play a similar role
for dynamics starting from the $|\mathbb{Z}_2\rangle$ state. The
minimum value of $\delta_3^n \equiv \delta^{n \,{\rm min}}_3$ as
well as the corresponding value of $h_3^{\rm min}$ decreases with
increasing $L$ as can be seen from Figs.\ \ref{3rdFSAplot} (b) and
\ref{3rdFSAplot} (c) respectively. We find that $\delta^{n\, {\rm
min}}_3 \to 0$ and $h^{\rm min} \to 0.29$ for sufficiently large but
finite $L$.

\begin{figure}
\includegraphics[width=\linewidth]{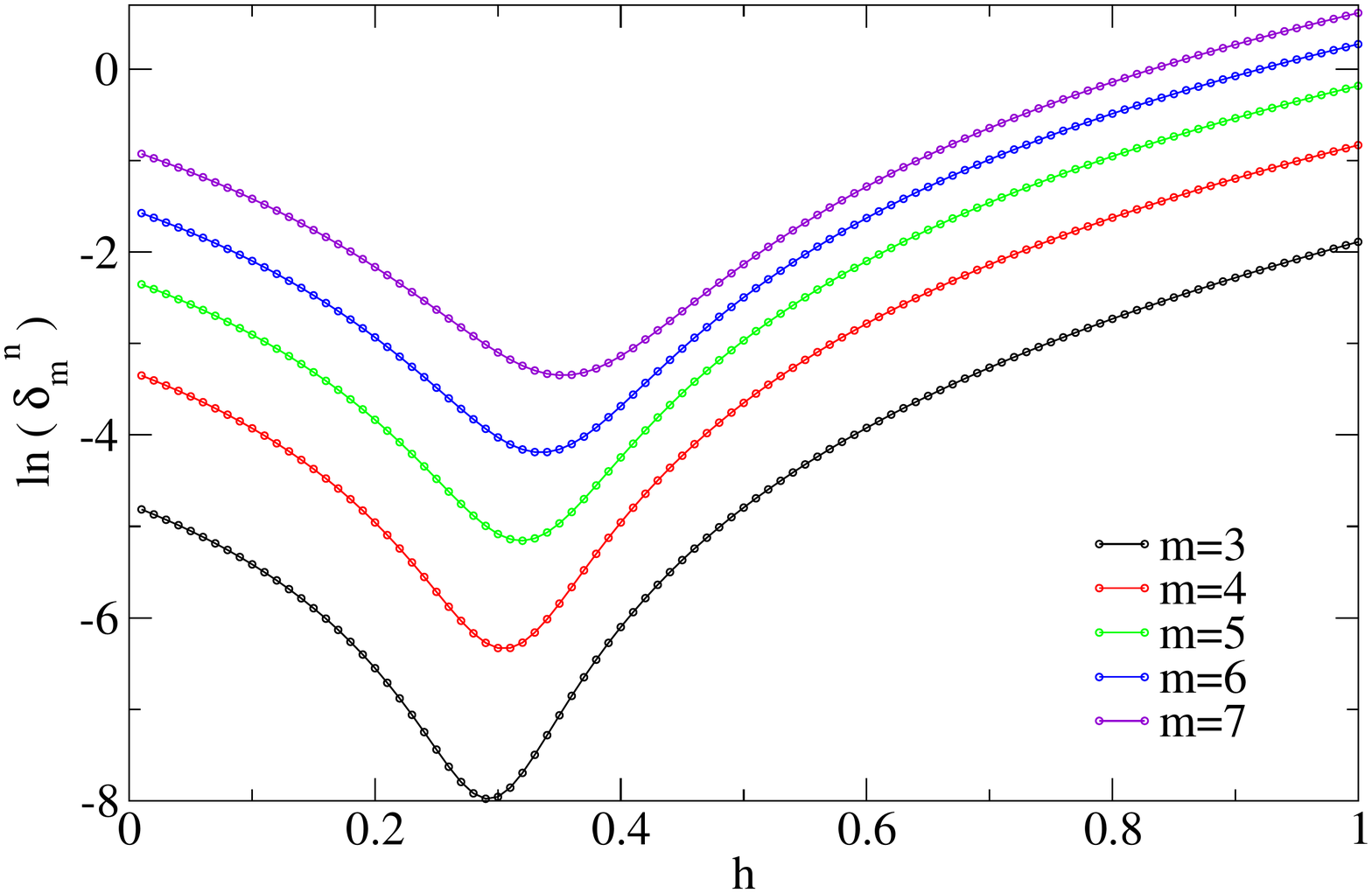}
\caption{Plot of $\delta_m^n$ for several FSA steps $m \le 7$ as a 
function of $h$ for $L=30$.} \label{error_n} \end{figure}

The above behavior of $\delta_3^n$ as a function of $h$ and $L$
suggests that the error in further FSA steps will play a crucial
role in determining the magnitude of oscillations in the dynamics of
the $|0\rangle$ state. However, a systematic analytic study of this seems
difficult. We therefore resort to numerical evaluation of these
errors denoted by $\delta_m^n$ for the $m^{\rm th}$ FSA step. The
result is shown in Fig.\ \ref{error_n} where we plot $\delta_m^n$ for
$m\le 7$ as a function of $h$ for $L=30$. We find that for all $h$,
$\delta_m^n$ is a monotonically increasing function of $m$. Moreover,
they display minima at different values of $h=h_m^{\rm min}$ which
are close to $h_3^{\rm min}=0.31$.

\begin{figure}
\includegraphics[width=\linewidth]{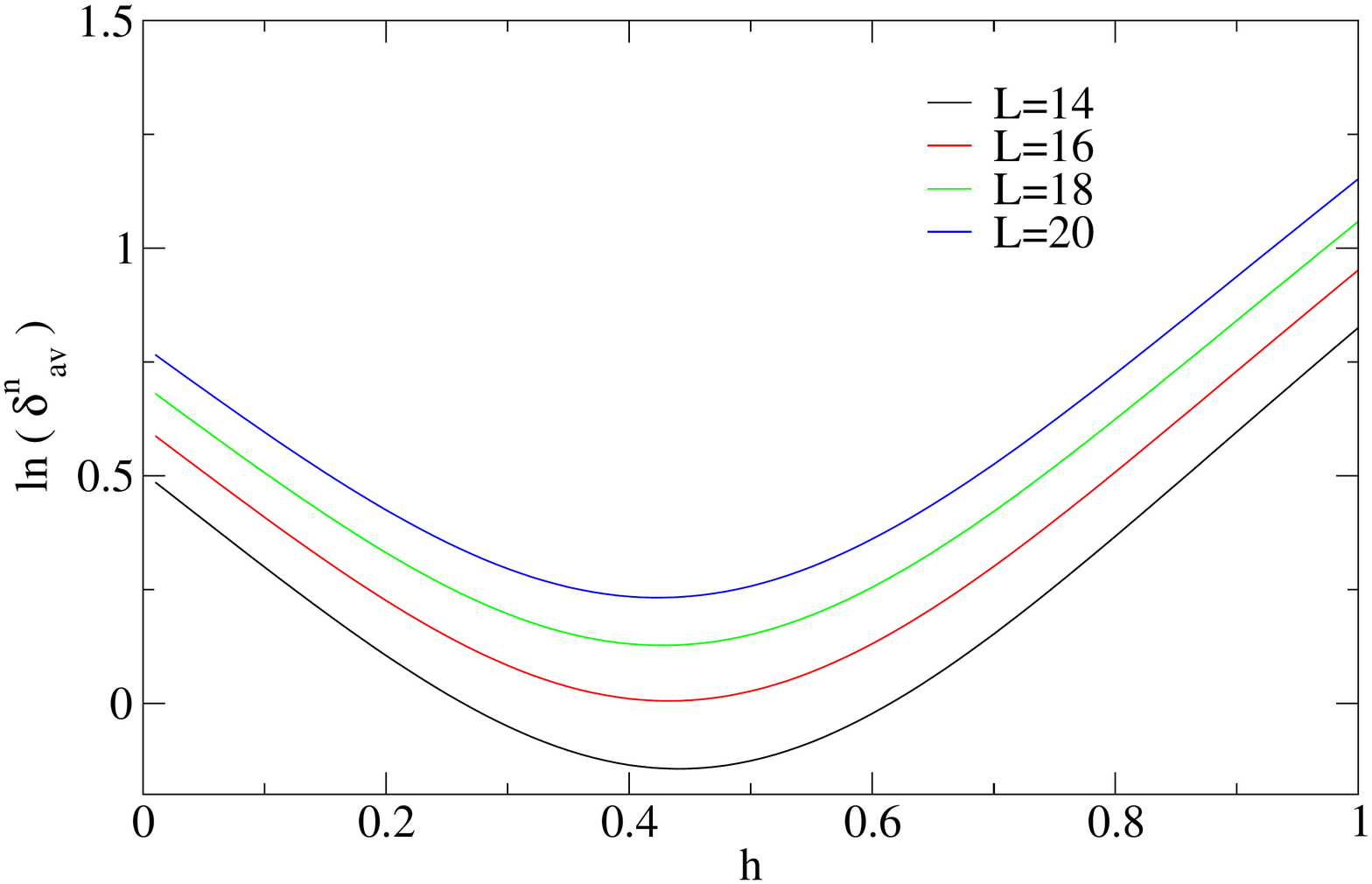}
\caption{Plot of $\ln \delta^n_{\rm av}$ as a function of $h$ for $L=14, 16, 18, 20$.}
\label{error_L} \end{figure}

In Fig.\ \ref{error_L} we plot $\ln \delta^n_{\rm av}$ as a function of $h$
for several representative values of $L$. Here, the average error
is defined as $\delta^n_{\rm av}=\sum_{m=3}^{L/2} \delta^n_m/((L/2)-2)$.
The plot indicates that tuning of $h$ can minimize the average
(and hence the total) error. However, the average error is a
monotically increasing function of $L$. This indicates decrease of
DE (and hence the oscillaion amplitude) value with $L$
in the superthermal phase as mentioned in the main text.

Next, we explore the possibility of maximizing the scar-induced
oscillations in these systems by allowing higher spin terms
\cite{cai}. These terms have support over $(2m+3)$ sites for a
$(2m+1)$-spin term. Here, we shall concentrate on the lowest such
term so that the Hamiltonian is
\begin{eqnarray} H_2 &=& H_1[h_3] + h_5 \sum_j (\tilde \sigma_{j-1}^+ \tilde
\sigma_{j+1}^+ \tilde\sigma^-_{j+2} \tilde\sigma_j^- \tilde\sigma_{j-2}^- +
      {\rm H.c.}),\nonumber\\
      \label{effectiveH}
\end{eqnarray}
such that the five-spin term has support over seven consecutive
sites. We note that it is experimentally challenging to generate
such terms with high enough amplitude; however, they are
automatically generated, albeit with lower strength, in our driven
system for the periodic protocol studied in this work. The aim of
our analysis here is to demonstrate that these terms indeed lead to
oscillations starting from the $|0\rangle$ state and that such
oscillations can be maximized by tuning their strength.

\begin{figure}
\includegraphics[width=\linewidth]{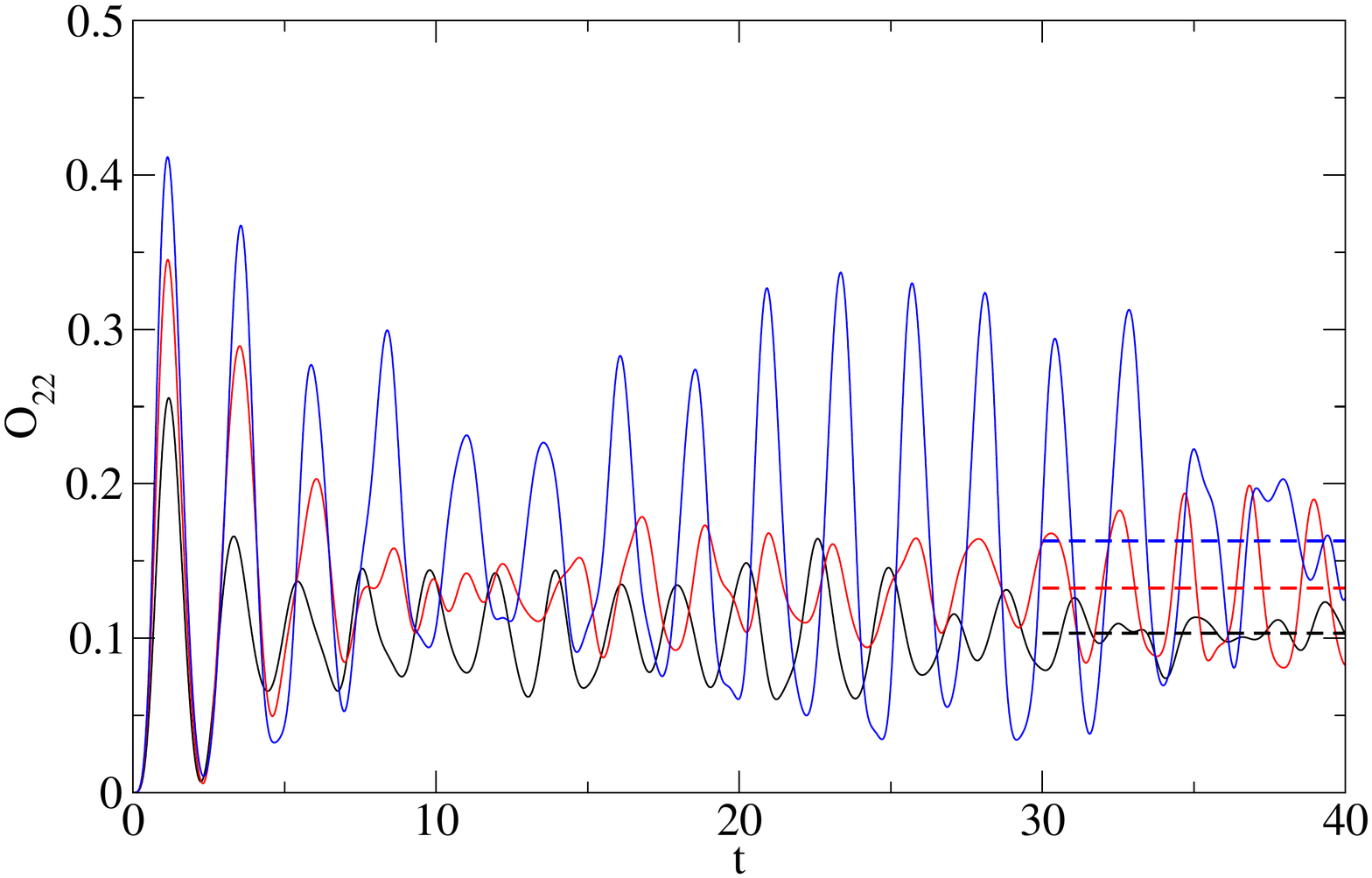}
\caption{Plots of $O_{22}$ as a function of $t$
  starting from $|0\rangle$ with the bare 
PXP Hamiltonian (black solid line), $H_1$ with $h=0.31$ (red solid line), and 
$H_2$ with $\vec h= (0.43, 0.28)$ (blue solid line). The corresponding diagonal
ensemble values are denoted by the dashed lines. For all plots, the system size equals $L=14$ and $t$ is measured in units of $w^{-1}$}. \label{osc} \end{figure}

To this end, we now repeat our analysis detailed out earlier using
$H_2$. Since our aim is to maximize the oscillation amplitude of
$O_{22} (n)$, we first numerically find the combination $\vec
h^{\rm max}= (h_3^{\rm max},h_5^{\rm max})$ which maximizes the DE
value of $O_{22}$; these values maximize the oscillation amplitude
as shown in Fig.\ \ref{osc} for $L=14$ where the initial state
$|0\rangle$ is propagated in time using $H_2$. We find that $\vec h^{\rm
  max}=(0.43,0,28)$. From Fig.\ \ref{osc}, it is clear that the
  oscillations in $O_{22}$ decrease when the $h_5$ term is set to
  zero and $h_3$ is
  then set to its optimum value, while the bare PXP Hamiltonian (setting
  both $h_3$ and $h_5$ to be zero) gives still weaker oscillations.

\begin{figure}
\includegraphics[width=\linewidth]{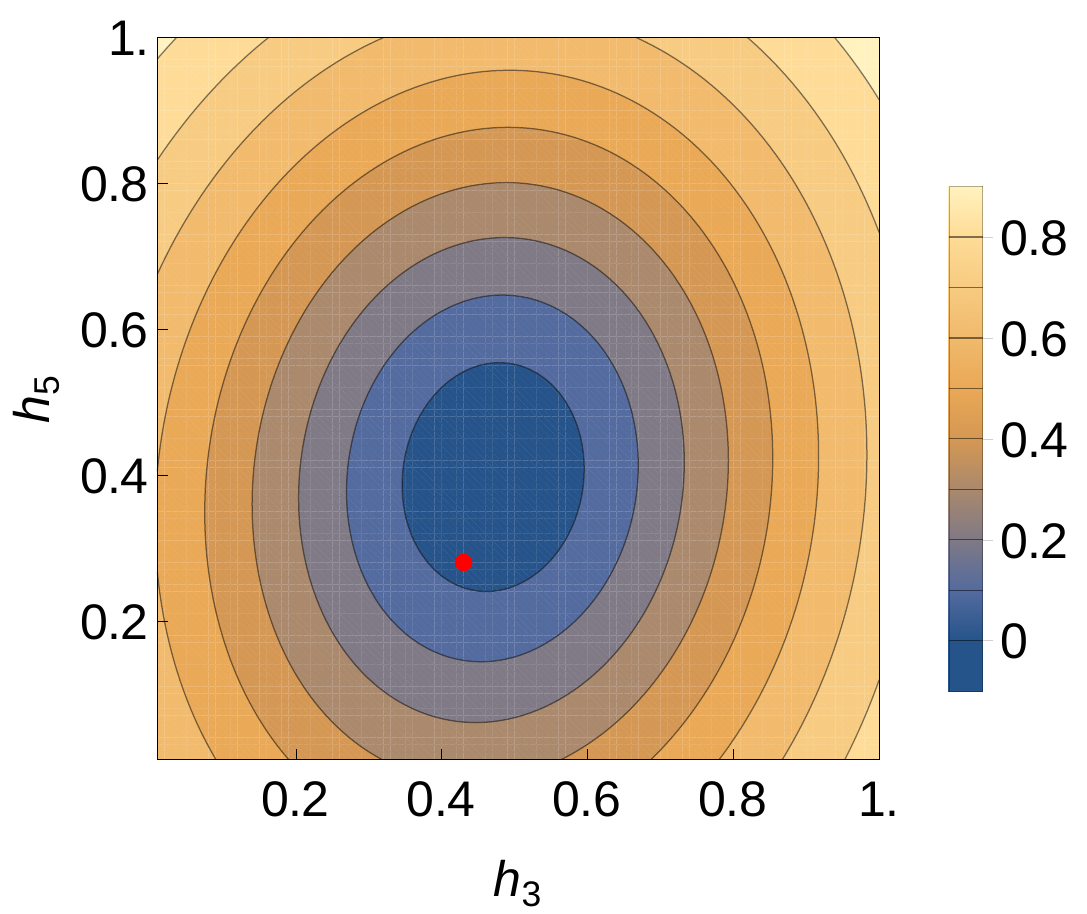}
\caption{Plot of $\ln \delta_{\rm av}^n$ as a function of $h_3$ and $h_5$ for
$L=14$. The red dot indicates the position of $\vec h^{\rm max}$.}
\label{comp} \end{figure}

Next, we carry out the FSA analysis using $H_2$ and starting from
the $|0\rangle$ state. Here we find that most of the features of
errors generated in different FSA steps mimics our earlier analysis.
In particular, we find that the FSA error at the $m^{\rm th}$ step
$\delta_m$ is minimized for different values of $\vec h$; there is
no value of $\vec h$ which minimizes all FSA errors. We therefore
choose to minimize $\delta_{\rm av}^n$. 
A plot of $\ln \delta_{\rm av}^n $ as a function of $h_3$ and $h_5$
is shown in Fig.\ \ref{comp}; this yields $\vec h^{\rm min}=
(0.45,0.4)$. The red dot in Fig.\ \ref{comp} indicates $\vec h^{\rm
max}$. We note that $\vec h_3^{\rm min}$ is close to $\vec h_3^{\rm
max}$ obtained earlier. However, the value $\vec h^{\rm min}$
depends on our choice of minimization parameter; for example,
minimization of errors of a specific step (say $\delta_5^n$ or
$\delta_6^n$) or the geometric mean of errors $(\prod_{m=3,7}
\delta_m^n)^{1/5}$ would leads to $\vec h^{\rm min}$ almost
identical to $\vec h^{\rm max}$. We leave a more detailed analysis
of this issue as a subject of future work.

\end{document}